\documentclass{aa}

\usepackage{graphicx}
\usepackage{txfonts}
\usepackage{hyperref}
\usepackage{orcidlink}

\begin{document}

\title{Ultraluminous X-ray sources in the first eROSITA survey}
\subtitle{I. Candidate catalogs}

\author{
  P.~Weber\inst{\ref{inst:remeis}}
  \and T.~P.~Roberts\inst{\ref{inst:durham}}
  \and S.~H\"ammerich\inst{\ref{inst:remeis}}
  \and E.~Kyritsis\inst{\ref{inst:heraklion},\ref{inst:forth},\ref{inst:mpe}}
  \and A.~Zezas\inst{\ref{inst:heraklion},\ref{inst:forth}}
  \and M.~G.~F.~Mayer\inst{\ref{inst:remeis}}
  \and A.~Zainab\inst{\ref{inst:remeis}}
  \and I.~Kreykenbohm\inst{\ref{inst:remeis}}
  \and A.~Merloni\inst{\ref{inst:mpe}}
  \and M.~Sasaki\inst{\ref{inst:remeis}}
  \and A.~Schwope\inst{\ref{inst:aip}}
  \and M.~Middleton\inst{\ref{inst:southampton}}
  \and A.~Basu-Zych\inst{\ref{inst:goddard},\ref{inst:maryland}}
  \and A.~Hornschemeier\inst{\ref{inst:goddard},\ref{inst:hopkins}}
  \and N.~Vulic\inst{\ref{inst:eureka},\ref{inst:goddard},\ref{inst:maryland},\ref{inst:greenbelt}}
  \and M.~Salvato\inst{\ref{inst:mpe}}
  \and N.~Webb\inst{\ref{inst:irap}}
  \and H.~Tranin\inst{\ref{inst:cosmos},\ref{inst:barcelona}}
  \and N.~Schettino\inst{\ref{inst:remeis}}
  \and C.~Kirsch\inst{\ref{inst:remeis}}
  \and S.~Saeedi\inst{\ref{inst:remeis}}
  \and F.~Zangrandi\inst{\ref{inst:remeis}}
  \and M.~Lorenz\inst{\ref{inst:remeis}}
  \and L.~Dauner\inst{\ref{inst:remeis}}
  \and T.~Dauser\inst{\ref{inst:remeis}}
  \and F.~Haberl\inst{\ref{inst:mpe}}
  \and C.~Maitra\inst{\ref{inst:iucaa},\ref{inst:mpe}}
  \and Z.~Igo\inst{\ref{inst:mpe}}
  \and A.~Santangelo\inst{\ref{inst:tuebingen}}
  \and L.~Ducci\inst{\ref{inst:tuebingen}}
  \and J.~Wilms\inst{\ref{inst:remeis}}
}

\institute{\label{inst:remeis}Dr.\ Karl Remeis-Sternwarte and Erlangen Centre for Astroparticle Physics, Friedrich-Alexander Universit\"at Erlangen-N\"urnberg, Sternwartstr.~7, 96049 Bamberg, Germany\\
    \email{philipp.ph.weber@fau.de}
\and
\label{inst:durham}Centre for Extragalactic Astronomy \& Dept of Physics, Durham University, South Road, Durham DH1 3LE, UK
\and
\label{inst:forth}Institute of Astrophysics, Foundation for Research and Technology-Hellas, GR 71110 Heraklion, Greece
\and
\label{inst:heraklion}Physics Department, \& Institute of Theoretical and Computational Physics, University of Crete, GR 71003, Heraklion, Greece
\and
\label{inst:aip}Leibniz-Institut für Astrophysik Potsdam, An der Sternwarte~16, 14482 Potsdam, Germany
\and
\label{inst:mpe}Max-Planck-Institut für Extraterrestrische Physik, Gießenbachstraße 1, 85748 Garching, Germany
\and
\label{inst:southampton}School of Physics and Astronomy, University of Southampton, Highfield, Southampton SO17 1BJ, UK
\and
\label{inst:eureka}Eureka Scientific, Inc., 2452 Delmer St., Suite 100, Oakland, CA 94602-3017, USA
\and
\label{inst:goddard}NASA Goddard Space Flight Center, Code 662, Greenbelt, MD 20771, USA
\and
\label{inst:maryland}Center for Space Science and Technology, U. of Maryland Baltimore County, 1000 Hilltop Circle, Baltimore, MD 21250, USA
\and
\label{inst:hopkins}Department of Physics and Astronomy, Johns Hopkins University, 3400 N. Charles Street, Baltimore, MD 21218, USA
\and
\label{inst:greenbelt}
Center for Research and Exploration in Space Science and Technology, NASA/GSFC, Greenbelt, MD 20771, USA
\and
\label{inst:irap}
IRAP, CNRS, Universit\'e de Toulouse, CNES, 9 Avenue du Colonel Roche, 31028 Toulouse, France
\and
\label{inst:cosmos}
Institut de Ci\`encies del Cosmos (ICCUB), Universitat de Barcelona (UB), c. Martí i Franqu\`es, 1, 08028, Barcelona, Spain
\and
\label{inst:barcelona}Dep. de F\'isica Qu\`antica i Astrof\'isica (FQA), Universitat de Barcelona (UB), c. Mart\' i Franqu\`es, 1, 08028, Barcelona, Spain
\and
\label{inst:tuebingen}Institut für Astronomie und Astrophysik, Universit\"at T\"ubingen, Sand 1, D-72076 T\"ubingen, Germany
\and
\label{inst:iucaa}{Inter University Centre for Astronomy \& Astophysics (IUCAA), Ganeshkhind, Pune 411007, India}
}

\date{Received XXXX; accepted XXXX}

  \abstract
  {
    Ultraluminous X-ray sources (ULXs) are luminous non-nuclear X-ray point
    sources embedded in galaxies. Their luminosities exceeding
    $10^{39}\,\mathrm{erg}\,\mathrm{s}^{-1}$ potentially require sub-Eddington
    accretion by objects with masses from ${\sim}10^2\,M_{\odot}$ to
    ${\sim}10^4\,M_{\odot}$ or supercritical accretion beyond the Eddington
    limit by stellar mass black holes or neutron stars. As such, ULXs are
    excellent laboratories for extreme accretion flows in X-ray binaries and, in
    some cases, potential candidates for intermediate-mass black holes. The
    first catalog by the eROSITA instrument on board the
    \textsl{Spectrum-Roentgen-Gamma} mission provides the most complete X-ray
    view of the whole of the western Galactic hemisphere and is therefore an
    excellent resource in which to find new ULXs.
  }
  {
    We identified ULXs by searching for X-ray counterparts in the ${\sim}10^5$
    local galaxies of the \mbox{\textsl{HECATE}} catalog in the footprint of the
    first eROSITA all-sky survey. The catalog is unaffected by X-ray
    selection biases. We characterized the completeness of the sample and its contamination by
    unidentified active galactic nuclei (AGN) and determined the influence of
    source confusion. The catalog provides the basis for ULX population studies
    and follow-up observations that can be used to perform a more detailed investigation of
    individual sources and variability.
  }
  {
    We searched for X-ray point sources exceeding the ULX threshold in
    \mbox{\textsl{HECATE}} galaxies while excluding the nuclear regions of the galaxies to avoid the
    selection of AGN. We removed known contaminants such as supernovae, stars,
    and AGN, and we manually vetted all candidates to further remove potential
    contaminants. We derived the X-ray selection function and fraction of
    unidentified AGN interlopers using a simulation with subsequent source
    detection. By comparing our sample to other ULX catalogs, we identified
    previously unknown candidates and those not detected although expected from
    their flux, indicating source intrinsic variability.
  }
  {
    We present two samples of ULX candidates. The main sample consists of 90
    sources with highly confident X-ray detections, 53 of which are identified
    for the first time. This sample is complete to a distance of ${\sim}7\,\mathrm{Mpc}$,
    contains at most 29\% of unknown background AGN, and is mostly unaffected by
    source confusion. Several candidates are newly identified even though their
    locations have been observed by other instruments before, and some
    previously reported candidates we expected to identify are absent from
    our sample, demonstrating the transient behavior of ULXs. Based on our list
    of less confident detections, we also provide an extended catalog of 260
    sources identified as potential candidates for further ULX identifications. Of these, 245 are identified as potential candidates for the first time.
  }
  {}
\keywords{}
\maketitle
\nolinenumbers
\section{Introduction}
Ultraluminous X-ray sources (ULXs) are defined as non-nuclear sources in nearby
galaxies with luminosities of ${\ge} 10^{39}\,\mathrm{erg}\,\mathrm{s}^{-1}$, that
is, larger than the Eddington luminosity of a ${\sim}10\,M_\odot$ black hole (BH) or
neutron star \citep[NS; see, e.g.,][for reviews]{kaaret2017,king2023}. While
sources later described as ULXs were already discovered by the first imaging
X-ray telescope, \textit{Einstein} \citep[e.g.,][]{fabbiano1987}, it is only 
in recent years that we have conclusively demonstrated that some of these
objects most likely constitute the extreme end of the ``normal'' X-ray binary
(XRB) population, given their similar hardness-ratio distributions
\citep{earnshaw2019,bernadich2022,tranin2024}. This is mainly due to the
detection of pulsations from several ULXs, indicating that they must harbor NS
accretors that far exceed the Eddington limit \citep[pulsating ULXs, or PULXs,
e.g.,][]{bachetti2014,israel2017a}. Most ULXs for which sufficient X-ray data
are available show a common range of X-ray spectral shapes and variability
properties that are unexpected from the ``classic'' sub-Eddington accretors in
our own Galaxy \citep[e.g.,][]{gladstone2009,sutton2013,middleton2015}, characteristics that are
suggestive of super-critical accretion
\citep[e.g.,][]{beloborodov1998,watarai2001,poutanen2007}. This super-critical
accretion scenario is also supported by the presence of wind signatures in X-ray
spectra \citep{middleton2014,pinto2016,kosec2021}. The similarities in spectral shape and
variability between PULXs and other well-studied ULXs indicate that NSs could
underlie a sizable part, if not the majority, of ULXs.
\citep[e.g.,][]{pintore2017,walton2018,gurpide2021}. The entire population may be
heterogeneous, however, with indications for intermediate-mass black holes (IMBH;
$M_{\mathrm{BH}} \sim 10^2$--$10^4\,M_\odot$) in a few very luminous objects
\citep[e.g.,][]{webb2012,mezcua2015}.

Catalogs have played a key role in improving our understanding of the ULX
population. The first ULX catalogs were by-products of the search for X-rays
from low-luminosity active galactic nuclei (AGN) with \textsl{ROSAT}
\citep{colbert1999,roberts2000} and included a few tens of sources. These were
then followed by systematic searches for ULXs in the \textsl{ROSAT} High Resolution Imager archives
that revealed around a hundred candidates \citep{colbert2002,liu2005}.
The number of ULX candidates has grown rapidly since then, with catalogs based
on several iterations of the \textsl{XMM-Newton} serendipitous source catalog
\citep{walton2011a,earnshaw2019,bernadich2022} or derived from the
\textsl{Chandra} data archives \citep{swartz2004,liu2011,kovlakas2020} revealing
hundreds of candidates. Currently, we now know of ${\sim}2000$ ULX
candidates, on the basis of meta-catalogs that merge the detections of ULXs in
the serendipitous source catalogs of \textsl{XMM-Newton}, \textsl{Chandra}, and
\textsl{Swift} \citep{walton2022,tranin2024}.

The ULX catalogs have provided insights into
the population demographics of ULXs. Early work revealed ULXs in both
elliptical and spiral galaxies, with an apparent correlation between the number
of ULXs and the star formation rate (SFR) of individual spiral galaxies
\citep{swartz2004,liu2005}. As the number of sources grew, the
incidence of ULXs per unit mass was found to be higher for lower-mass spiral galaxies
\citep{walton2011a}. The number of ULXs in a star-forming
galaxy is known to scale with the SFR and stellar mass ($M_*$; \citep{kovlakas2020}),
which is consistent with population synthesis models that take into account stellar ages
and metallicity. Roughly one in three galaxies host at least one ULX
\citep[see also][]{earnshaw2019}. Catalogs have also provided target
lists for further study, which is particularly important for samples of
rare and interesting objects. For example, there has been a study following up
on sub-samples of the most luminous and/or brightest ULXs in the search for IMBH
candidates \citep{sutton2012}, which included the first investigation of the
most luminous PULX, NGC~5907~ULX \citep{roberts2023}. 
The catalogs have also been used for variability studies in the search
for PULX candidates \citep{earnshaw2018} or to constrain the possible NS
fraction of the ULX population \citep{khan2022}.

All existing ULX catalogs have strong biases originating from the serendipitous
source samples they are based on. These samples are mostly based on pointed
observations that do not have ULX searches as their primary goal. While many
catalogs contain ``complete'' sub-samples that, for instance, include all
candidates from galaxies for which all ULXs can be detected (excluding some
regions around the nucleus), the galaxies in which they reside are not selected
as representative in any way. The observations that underlie these searches are often biased toward systems
that are interesting for other reasons, such as an active nucleus, galaxy
interaction, or other sources of high SFR or mass. Therefore, ULX
population studies can benefit immensely from a new sample of
candidates based on completely unbiased X-ray observations of galaxies.

The catalog of the first eROSITA X-ray all sky survey (eRASS1) is the deepest
and most uniform X-ray survey of the western Galactic hemisphere, and it enables
searches for ULXs in an unbiased way. We present the first catalog of
ULX-candidates from eRASS1 that is deep enough to find all active ULXs out to
${\sim}7\,\mathrm{Mpc}$. In Sect.~\ref{sec:catalog} we present the eRASS source catalog and our
identification of ULX candidates,followed by a description of basic sample properties,
such as completeness and residual contamination (Sect.~\ref{sec:sample}). In
Sect.~\ref{sec:discussion} we discuss the limitations of the catalog and its
main differences compared to other existing catalogs generated from pointed
observations, and this is followed by the exploration of possible source intrinsic
variability in Sect.~\ref{sec:variability}. We summarize our results in
Sect.~\ref{sec:summary} and provide an outlook on population studies based on
our catalog in an accompanying paper (Weber et al., 2026b, hereafter paper~II).

\section{Identifying ULX in the eROSITA all sky survey}\label{sec:catalog}
In this section we describe the data used for the creation of the catalog and
their extraction as well as the steps undertaken to derive the parameters necessary
for the identification of ULX candidates. We start with descriptions of the
eROSITA survey (Sect.~\ref{sec:erosurvey}) and the \mbox{\textsl{HECATE}}
catalog (Sect.~\ref{sec:hecate}), followed by a detailed explanation of the ULX
candidate selection (Sect.~\ref{sec:identification}).

\subsection{The eROSITA all sky survey}\label{sec:erosurvey}
eROSITA is the soft instrument on board the \textsl{Spectrum-Roentgen-Gamma}
mission \citep{sunyaev2021}. Launched on 2019 July 13 to the second
Lagrange point of the Sun--Earth system, it consists of seven co-aligned Wolter
type~I telescope modules (TMs) combined with X-ray charge coupled devices
sensitive from $\sim$0.2\,keV to $\sim$8.0\,keV, providing a field of
view of $\sim$1\,degree \citep{predehl2021}.

From 2019 December until 2022 February, \textsl{Spectrum-Roentgen-Gamma}
performed an all-sky survey by rotating around its Earth-pointing axis with a
period of 4\,h. The passage of each location on the sky through the field of
view takes ${\sim}40\,\mathrm{s}$. The survey poles are roughly aligned with the
ecliptic poles. Positions at the ecliptic equator are covered by six subsequent
scans, while the poles are covered during each scan, resulting in exposure times
from ${\sim}0.2\,\mathrm{ks}$ to ${\sim}4\,\mathrm{ks}$ respectively
\citep{predehl2021}. As eROSITA moves around the Sun, the entire sky is covered
once every 6\,months, or one eRASS. Access to eROSITA data is split at the
Galactic longitude of Sgr~A$^*$, $l_{II} \simeq 0^\circ$. The German eROSITA
consortium is responsible for the western Galactic hemisphere. Data and
higher-level data products taken in this region during eRASS1 were made publicly
available on 2024 January 31
\citep{merloni2024}\footnote{\url{https://erosita.mpe.mpg.de/dr1/}}. Our work is
based on the eRASS1 main catalog of 930203 sources detected in the 0.2--2.3\,keV
band, which lists general source properties such as count rates and fluxes. In
this work we consider the astrometrically corrected positions, unless stated
otherwise.

\subsection{The \mbox{\textsl{HECATE}} galaxy sample}\label{sec:hecate}
To identify ULXs, we needed to determine which point sources in the
eRASS1 catalog are colocated with known galaxies by comparing their positions with the $D_{25}$ regions. While \citet{walton2022} used the
\mbox{\textsl{HyperLEDA}} catalog \citep{makarov2014}, \citet{tranin2024} used the \mbox{\textsl{GLADE}} catalog
\citep{dalya2018}, which extends \mbox{\textsl{HyperLEDA}} with information from other sources.
Following the approach of \citet{bernadich2022}, we chose the Heraklion Extragalactic Catalogue (\mbox{\textsl{HECATE}}) compiled by
\citet{kovlakas2021} based on the \mbox{\textsl{HyperLEDA}} catalog as the main source of
information on galaxies in the eRASS1 footprint.
\mbox{\textsl{HECATE}} contains the positions and
distances of 204733 galaxies, with additional information on their intrinsic
properties, including morphological types, 
luminosities, star formation rates, and metallicities. It also supplies the
apparent geometric shape of the galaxies in the form of $D_{25}$ regions, which
are ellipses corresponding to the $25\,\mathrm{mag}\,\mathrm{arcsec}^{-2}$ $B$~band isophote, parametrized with semi-major and semi-minor axes, $r_1$ and $r_2$, and the position angle of $r_1$.
We based our work on a revised 
version of \mbox{\textsl{HECATE}} that includes improved values for
the host properties, most notably for our purposes their distances \citep{kyritsis2026}. Since
\mbox{\textsl{HECATE}} is based on \mbox{\textsl{HyperLEDA}}, which was assembled from a
variety of surveys with different footprints, it is not strictly possible to
reliably determine its completeness \citep{kovlakas2021}. For $B$~band
luminosities above ${\sim}10^{7.1}\,L_{\mathrm{B},\odot}$ the new version of
\mbox{\textsl{HECATE}} is complete within 10\,Mpc, and down to
${\sim}10^{10.7}\,L_{\mathrm{B},\odot}$ between ${\sim}40\,\mathrm{Mpc}$ and
${\sim}50\,\mathrm{Mpc}$ \citep{kyritsis2026}. Compared to \mbox{\textsl{GLADE}},
\mbox{\textsl{HECATE}} provides more robust distance estimates by accounting for
deviations from the Hubble flow. This is especially important given the
localized sensitivity to ULXs in the eRASS1. Additionally,
\mbox{\textsl{HECATE}} directly provides extent information in the form of
$D_{25}$ ellipses, star formation rates, stellar masses, and metallicities, all
of which can be used for further population studies (paper~II).

\subsection{ULX candidate identification and data reduction}
\label{sec:identification}

\begin{figure}
  \resizebox{\hsize}{!}{\includegraphics{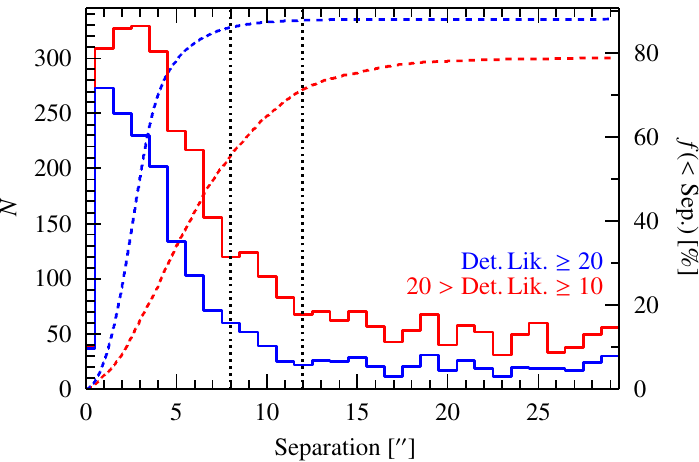}}
  \caption{
    Distribution of the angular separations between the cores of the \mbox{\textsl{HECATE}}
    $D_{25}$ regions and the closest eRASS1 source (solid, left axis), and the
    cumulative distribution of the discrepancy between the initial and detected
    source position determined from a simulation of eRASS1 (dashed, right axis).
    We show both sets for different detection likelihood intervals. Separations
    of $8''$ and $12''$ are marked. For both sets of eRASS1 sources, the wing of
    central matches extends beyond $8''$ to $12''$. For lower detection
    likelihoods, a sizeable fraction of simulated sources was detected beyond $8''$.
  }
  \label{fig:hecate_cores}
\end{figure}
To identify ULX candidates we considered all point sources in the eRASS1 catalog
located within the $D_{25}$ ellipse of \mbox{\textsl{HECATE}} galaxies. The
positional uncertainty of the X-ray detections can be ignored for this step
since it generally is small compared to the semimajor axes of the galaxies
(${\lesssim}5\%$). However, we ignored all sources without positional error, that
is, \verb|RA_LOWERR|, \verb|RA_UPERR|, \verb|DEC_LOWERR|, or \verb|DEC_UPERR| is
\verb|NaN|. This is necessary because the absence of these uncertainties
indicates a failed point spread function  fit at the source position, often caused by strong,
extended emission not associated with a point source, such as present in
Messier\,87. We also ignored sources where detection problems are indicated by at
least one of the quality flags \verb|FLAG_SP_SNR|, \verb|FLAG_SP_BPS|,
\verb|FLAG_SP_SCL|, \verb|FLAG_SP_LGA|, \verb|FLAG_SP_GC_CONS|,
\verb|FLAG_NO_RADEC_ERR|, \verb|FLAG_NO_CTS_ERR|, and \verb|FLAG_NO_EXT_ERR|
\citep[see][Sect.~5.2 and Table~5 for a description of the quality
flags]{merloni2024}. These quality criteria removed ${\sim}3.5\%$ of sources from
the eRASS1 catalog.

\begin{figure}
  \resizebox{0.48\hsize}{!}{\includegraphics{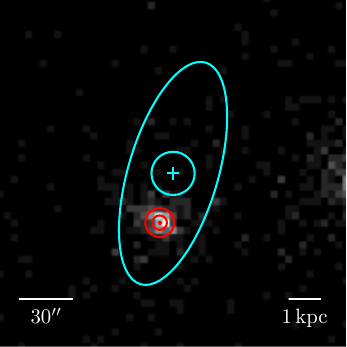}}\hfill
  \resizebox{0.48\hsize}{!}{\includegraphics{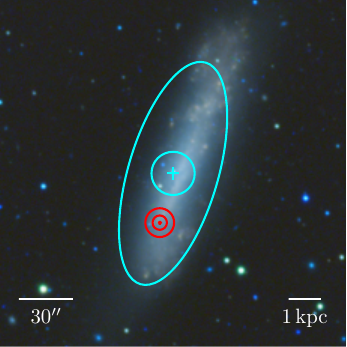}}\\
  \resizebox{0.48\hsize}{!}{\includegraphics{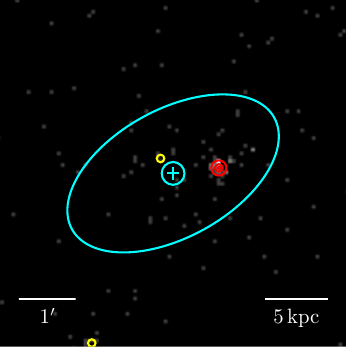}}\hfill
  \resizebox{0.48\hsize}{!}{\includegraphics{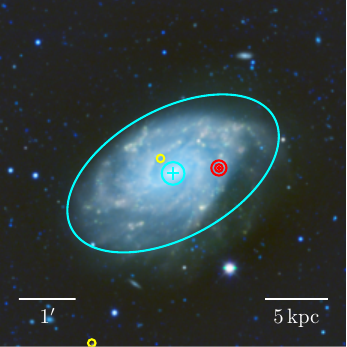}}\\
  \resizebox{0.48\hsize}{!}{\includegraphics{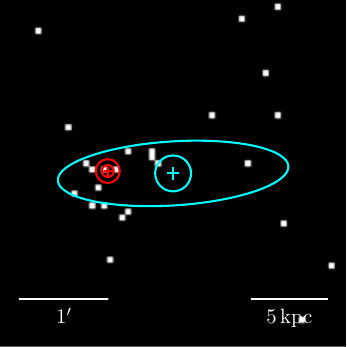}}\hfill
  \resizebox{0.48\hsize}{!}{\includegraphics{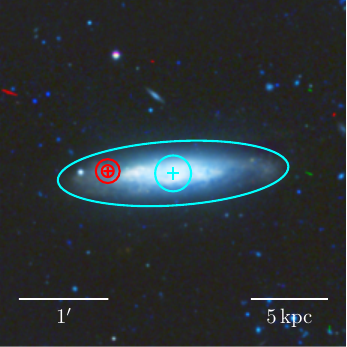}}
  \caption{X-ray and optical images of selected ULX candidates. The left column shows 0.2\,keV--5.0\,keV X-ray
    images; the right columns show optical images of the same
    area in the g, r, and z bands taken from DESI Legacy Imaging Survey DR8 \citep{dey2019}. The image scale is
    shown in the bottom left-hand corner, the projected extent in the frame of
    the galaxy in the bottom right-hand corner. The cyan ellipse is the $D_{25}$
    area as listed in \mbox{\textsl{HECATE}}. Its central position is denoted by a cross,
    and the $12''$ radius circular exclusion zone around the center is marked as a circle. The position of the
    X-ray source is marked with red circles of a $4''$ and $8''$ radius; other eRASS1 sources are indicated as yellow circles. The first
    two rows show typical examples for candidates in the main ULX sample
    (NGC~1824 in the first row, NGC~1255 in the second row). The third row shows
    a typical candidate of the extended, less secure sample in NGC~4129. }
  \label{fig:images}
\end{figure}
Low luminosity AGN (LLAGN) in the centers of galaxies have luminosities above the ULX
threshold and can be confused with ULXs. To remove such interlopers, we excluded
all sources identified too close to the nucleus of each galaxy. In
Fig.~\ref{fig:hecate_cores} we show the distribution of the distances of the
closest matches between the cores of the $D_{25}$ ellipses and the eRASS1
sources. Most matches are concentrated around ${\sim}3''$, but a wing of matches
extends to ${\sim}12''$. The most likely explanation for this distribution is a
combination of the positional uncertainty of LLAGN in the
center of some galaxies, together with the fact that the center of the $D_{25}$
ellipse does not necessarily correspond to the nucleus of a given galaxy since
it only optimizes the isophotes in the outer areas and does not relate to the
surface brightness in the core. Late-type galaxies especially can have complex
surface brightness distributions (see Fig.~\ref{fig:images}), which hinders
the determination of the engulfing $D_{25}$ ellipse and therefore the
center of the galaxy. Additionally, a putative AGN does not necessarily have to
coincide with the center of its host, which is particularly relevant for
dwarf galaxies \citep{reines2020,bellovary2021} and mergers \citep{barrows2016}.
In Fig.~\ref{fig:hecate_cores} we also show the positional deviation between the
assumed and detected positions of simulated eRASS1 sources (see
Sect.~\ref{sec:completeness} for details regarding the simulation) and find that
for detection likelihoods between 10 and 20 a significant portion of sources are
detected further than $8''$ from their simulated position but mostly within
$12''$. We thus ignored all sources closer than $12''$ to the core of their
respective $D_{25}$ ellipse. After this step, 10474 sources remain in 866
galaxies.

To derive source luminosities, we extracted spectra with \texttt{srctool}
version 1.76.2 from eSASS release 211214, patch level 4 \citep{brunner2022}. To
optimize the signal of the spectra, the extraction radii for the circular source
regions need to be large enough to encompass most source events but small enough
to minimize the inclusion of background events. In the standard data reduction
performed for eRASS1 \citep{brunner2022}, source extraction region sizes are
optimized to maximize the signal-to-noise ratio with \texttt{srctool} based on a
set of information from the source detection process that is not available for
our analysis.\footnote{The algorithm is described at
\url{https://erosita.mpe.mpg.de/dr1/eSASS4DR1/eSASS4DR1_tasks/srctool_doc.html}} We therefore obtained extracted products from the
first eROSITA data release\footnote{\url{https://erosita.mpe.mpg.de/dr1/}} for
all sources in the eRASS1 and found an empirical relation between the maximum
likelihood count rate \verb|ML_RATE_0| and the radii of the extraction regions,
$r$, of
\begin{equation}
  r = \mathrm{min}\left(200'',\mathrm{max}\left(85\farcs4 \cdot \verb|ML_RATE_0|^{0.28}, 23''\right)\right) .
\end{equation}
In this relation \verb|ML_RATE_0| is a proxy for the flux, capturing the
larger spread of photons across the sensor for brighter sources.
We extracted background spectra based on annuli centered on the source region
and outside of the source circle, excluding all eRASS1 point sources from the
background region, with
\begin{equation}
  r_{\mathrm{in}} = \mathrm{min}\left(350'', \mathrm{max}\left(182\farcs0 \cdot \verb|ML_RATE_0|^{0.24}, 54''\right) \right)
\end{equation}
for the inner radius and 
\begin{equation}
  r_{\mathrm{out}} = \mathrm{min}\left(2200'', \mathrm{max}\left(1063\farcs2 \cdot \verb|ML_RATE_0|^{0.28}, 280''\right) \right) 
\end{equation}
for the outer radius.

We derived the fluxes of all candidates by modeling their spectra. The spectra of
ULXs generally have complex shapes with multiple soft components and a high
energy cutoff, known to vary between
observations \citep{kaaret2017,fabrika2021,king2023}. At this stage in the
selection process the median number of detected photons per candidate was 11,
which rendered a spectral fit with multiple components unfeasible. We therefore
adopted a single absorbed power law with a slope of $\Gamma = 1.7$. Given their
location within other galaxies, the emission of ULXs is not only affected by
Galactic foreground absorption, but also absorption intrinsic to the host
galaxy. Typically, the intrinsic equivalent hydrogen column of ULXs is around
${\sim}10^{21}\,\mathrm{cm}^{-2}$ \citep{kaaret2017}. Given our available data,
fitting a model with varying $N_{\mathrm{H}}$ proved statistically challenging.
We therefore used a single absorption component with a fixed $N_{\mathrm{H}}$ of
$10^{21}\,\mathrm{cm}^{-2}$. We fitted the 0.2\,keV--5\,keV spectra for all TMs
except for TM5 and TM7, as for these TMs data are affected by spectral
distortion caused by a light leak in the sensor chamber \citep{predehl2021},
necessitating an increase of the lower bound to 1\,keV. This energy band is
narrower than the full energy range provided in the main source catalogs derived
from data by \textsl{Swift} \citep[2SXPS, ][]{evans2020}, \textsl{Chandra}
\citep[CSC2, ][]{evans2024}, or \textsl{XMM-Newton} \citep[4XMM, ][]{webb2020},
which have been used in other recent ULX compilations. Compared to the widest
available energy range of these instruments, from 0.2\,keV to 12.0\,keV by
\textsl{XMM-Newton}, the resulting flux is reduced by ${\sim}40\%$, assuming the
same spectral shape. For a slope of the ULX X-ray luminosity function
(XLF) of $\alpha = 1$ \citep{tranin2024} this reduced flux leads to an expected
decrease of ULX classifications by ${\sim}40\%$. The spectra of sources with a
detection likelihood of 10 typically contain ${\sim}10$ net counts with a
background contribution of ${\sim}2$ counts. We therefore did not 
extrapolate the flux to the energy bands of these other instruments as it
would have inflated its uncertainty. Unless indicated otherwise, all spectral fits
quoted here and in the following were performed on background subtracted spectra
with ISIS, version 1.6.2-51 \citep{houck2000}, and utilized \citet{cash1979}
statistics. We derived luminosities from the best-fit fluxes using the distances
of the host galaxies as listed in \mbox{\textsl{HECATE}}. Point sources were
retained as ULX candidates if their 0.2--5.0\,keV band luminosity exceeded
$10^{39}\,\mathrm{erg}\,\mathrm{s}^{-1}$, which left us with 838 ULX
candidates in 784 galaxies. Choosing a more shallow (steeper) slope with $\Gamma
= 1.5$ ($\Gamma = 2.0$) would have increased (decreased) the flux by ${\sim}10\%$, which
would have resulted in 843 (830) candidates at this stage.

\subsection{Sample cleaning}\label{sec:cleaning}
\begin{figure}
    \resizebox{\hsize}{!}{\includegraphics{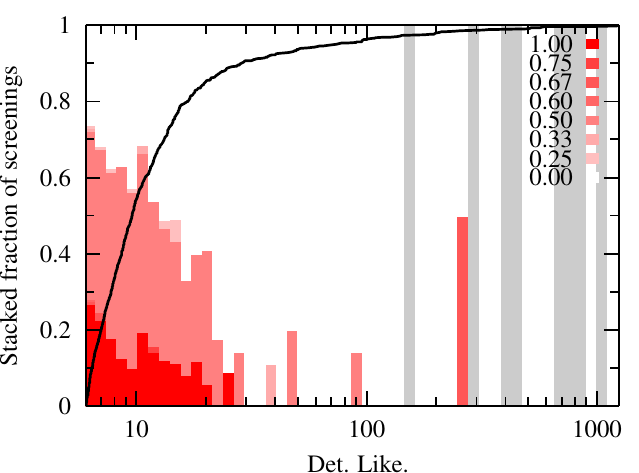}}
    \caption{
      Stacked bar diagram showing how often a candidate was flagged as a spurious
      detection compared to how often it was screened with respect to the
      detection likelihood. The opaqueness of the red bars (including the white
      areas) denote the relative number of times a source within each area was
      flagged as spurious. Sources were investigated between two and five times,
      leading to the fractions in the legend. The sizes of the bars correspond
      to the relative number of sources within the bin falling in the respective
      category. The black line in the foreground shows the cumulative
      distribution of the entire candidate sample. Gray areas denote empty bins.
      The peak at ${\sim}250$ is caused by one of two sources flagged within
      this bin.
      }
    \label{fig:screening_detlike}
\end{figure}
The eRASS1 catalog provides information about the confidence of individual
detections in the form of the detection likelihood, which was derived by fitting
a calibrated point spread function profile to the measured photon distribution
at the position of a
putative source \citep{merloni2024}. Generally, this likelihood scales with the
number of counts available for the detection of a source and therefore with its
flux and exposure. In order to define a reliable sample of candidates, we needed to
find a lower limit for this detection likelihood, above which spurious detections
due to individual background photons resembling a weak point source can be ruled
out. Given the complex environment in which the ULX candidates are located, such
as a background of diffuse emission from the host galaxy, it is not sufficient
to use purely statistical arguments at this step. To this end we conducted a
manual screening process during which each of the candidates was investigated by
at least two distinct and randomly assigned astronomers. Each person was
provided with X-ray and optical images from eROSITA, from the Digitized Sky
Survey \citep{lasker1996}, Panoramic Survey Telescope and Rapid Response System
\citep[Pan-STARRS]{chambers2016}, and the Dark Energy Survey
\citep[DES]{dey2019}, similar to those in Fig.~\ref{fig:images}. Additionally,
supplemental information including light curves of flux and luminosity for all
four eRASS surveys (determined from the current internal version of the eRASS:4
catalog), basic source parameters, derived parameters (flux, luminosity, upper
limits of flux, luminosity, and count rate), and various other information, such
as potential counterparts and past observations were available. Each person
could then assign a variety of flags to each candidate, indicating special
properties of the source (e.g., whether the source coincides with a clear point
source, is located outside of the galaxy, or is not a point source), and note
comments for further discussion. See Appendix~\ref{app:flags} for a complete
list of available flags in the screening process. We provide the results of the
screening process as columns in the published catalog, most notably
\verb|screen_flag_spurious| (see Appendix~\ref{app:columns}).
Figure~\ref{fig:screening_detlike} compares the detection likelihood of the
candidates with the frequency of how often they were flagged as spurious
detections. Based on these results, we consider sources with detection
likelihoods above 20 as members of our  main sample. Candidates with detection
likelihoods between 10 and 20 form part of an extended sample with pending
confirmation of being real X-ray detections. We ignored all other sources with
detection likelihoods of less than ten. Using these limits, we identified 111
candidates in 104 galaxies in the main sample and 268 candidates in 264 galaxies
in the extended sample.

\citet{seppi2022} determined the detectability of galaxy clusters and AGN with
eROSITA through a simulation with subsequent source detection. They find an
overall rate of ${\sim}1\%$ spurious AGN detections due to random background
fluctuations at a lower detection likelihood limit of 10. However, this number
was derived from sources isolated from other environments, specifically not
embedded in the diffuse emission of galaxies, which is the case for ULXs. The
elevated rate of diffuse emission increases the number of spurious
detections. Additionally, extended structure with low flux in the galaxy, such
as star-forming regions, cannot be
detected as such and registers as spurious point sources. Based on the manual
screening, we therefore found that a more conservative minimum detection likelihood
is necessary to keep our ULX samples clean from such interlopers.

To reduce the number of contaminants further, we performed an extensive search
for possible counterparts of a variety of other source types, utilizing
positional matching with a maximum source distance of $8''$. These
identifications are not mutually exclusive. Particularly, during the manual
screening process, a flag could be set that duplicates the result of the
automatic matching or contradicts it. Supernovae in other galaxies can exceed
luminosities of $10^{39}\,\mathrm{erg}\,\mathrm{s}^{-1}$. They occur over the
entire extent of the galaxy, resulting in an appearance similar to that of ULXs.
We therefore matched our candidates against the Open Supernova Catalog
\citep{guillochon2017} and queried the NASA/IPAC Extragalactic Database
\citep{Helou1991} for known supernovae, which identified 11 counterparts. Matching
against the AGN catalog of \citet{veron-cetty2010}, we identified three possible
contaminants. To identify foreground stars we matched the samples against the Gaia
DR3 catalog \citep{gaiacollaboration2023} to find counterparts with a
parallax larger than 0, as well as the Tycho-2 Catalogue \citep{hog2000}, and the
Yale Catalog of Bright Stars \citep{hoffleit1995}, yielding three foreground
stars in total. We also removed one of the five ``fake'' ULXs identified by
\citet{gutierrez2013} from the main sample, and one object identified as
background AGN \citep{sutton2015}. One further candidate was flagged manually
during visual inspection as background AGN and seven as foreground stars. Four
additional candidates were identified in galaxies with $D_{25}$ regions that do
not resemble the optical shape of the galaxy. We supply a comprehensive list of
all identified contaminants in order to help future ULX searches. 

After the removal of these identified contaminants, the main and extended samples
consist of 90 candidates in 85 galaxies and 260 candidates in 256 galaxies
respectively. Five hosting galaxies in the main sample and four hosting galaxies
in the extended sample contain two candidates. Of all the candidates, 37 in our main sample
and 15 in the extended sample were previously listed by \citet{kovlakas2020},
\citet{walton2022}, \citet{bernadich2022}, or \citet{tranin2024}. Therefore, there
are 53 previously unknown ULX candidates in the main sample and 245 in the extended
sample.
We provide the samples and identified contaminants on
the DR1
page.\footnote{\url{https://erosita.mpe.mpg.de/dr1/AllSkySurveyData_dr1/Catalogues_dr1/}}

\section{The eROSITA ULX candidate sample}\label{sec:sample}
In this section we describe the properties of the ULX candidate sample. We start
by deriving the selection function through a simulation based on which we
calculate the completeness of the sample (Sect.~\ref{sec:completeness}). Using this result, we estimate
in Sect.~\ref{sec:contamination} how the sample is contaminated by background AGN.

\subsection{Completeness}
\label{sec:completeness}
\begin{figure}
  \resizebox{\hsize}{!}{\includegraphics{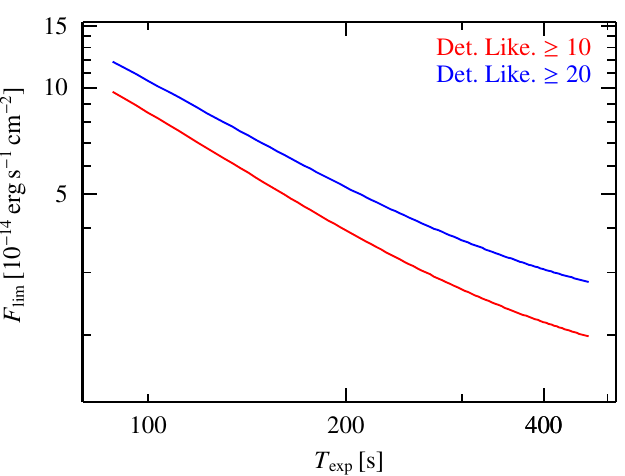}}
  \caption{Simulated limiting flux for a detection with at least 95\% probability
  depending on the exposure time at different detection likelihood thresholds.}
  \label{fig:sensitivity_exposure}
\end{figure}

\begin{table}
  \centering
  \caption{Parameters for Eq.~\ref{eq:sensitivity} to give a lower flux
  limit for a detection with a probability of 95\% at different detection
  likelihoods.}
  \label{tab:sensitivity_parameters}
  \begin{tabular}{lcccc}
  \hline
    \hline
    Det.\ Like.& $a$ & $\beta$ & $t_0$ & $F_0$\\
    \hline
    ${\ge}10$  &   $2.2$   & $2.1$ & $-55$ & $1.4$ \\
    ${\ge}20$  &   $2.2$   & $2.0$ & $-60$ & $2.1$ \\
    \hline
  \end{tabular}
  \tablefoot{Units are: $t_0$: s,  $a$:
  $10^{-9}\,\mathrm{erg}\,\mathrm{s}^{-1}\,\mathrm{cm}^{-2}$, $F_0$:
  $10^{-14}\,\mathrm{erg}\,\mathrm{s}^{-1}\,\mathrm{cm}^{-2}$.
  The fits are based on values determined from a cumulative
  distribution and therefore lack uncertainties.
  }
\end{table}

The effective exposure of individual eROSITA sources is a strong function of
their position, both due to the survey strategy and due to strong vignetting
(Sect.~\ref{sec:erosurvey}), implying a complex source detection behavior. To
quantify this performance, we adopted the approach by \citet{seppi2022} for our
selected detection likelihood limits and the spectral shape assumed for our ULX
candidates. We used the SIXTE simulation package \citep{dauser2019} and the
eROSITA attitude for the western Galactic hemisphere to generate simulated
eROSITA events of $10^5$ randomly distributed sample sources. The fluxes
of the sources were logarithmically uniformly distributed between
$10^{-16}\,\mathrm{erg}\,\mathrm{s}^{-1}\,\mathrm{cm}^{-2}$ and
$10^{-11}\,\mathrm{erg}\,\mathrm{s}^{-1}\,\mathrm{cm}^{-2}$ in the 0.2--2.3\,keV
band with an absorbed power law spectrum ($\Gamma = 1.7$, $N_{\mathrm{H}} =
10^{21}\,\mathrm{cm}^{-2}$). This flux range ensures the coverage of the full
eROSITA sensitivity range. We then processed these simulated event data with
eSASS, following the same steps that were used for the creation of the eRASS1
catalog \citep{merloni2024}. By comparing the detected simulated sources with
the input catalog, we determined the detection probability as a function of flux
and exposure, and the number of false positives, that is, the sources detected by
eSASS at locations on the sky where no sample source is located. Specifically,
we considered a source to be detected if it is within $8''$ of an input source.
Using the detection likelihood thresholds 10 and 20, we could then determine the
minimum flux above which at least 95\% of the input sources were detected as a
function of exposure time, $T_\mathrm{exp}$
(Fig.~\ref{fig:sensitivity_exposure}).  We find that this limiting 0.2--2.3\,keV
flux can be described by
\begin{equation}  \label{eq:sensitivity}
  F_\mathrm{lim}(T_\mathrm{exp}) = a \cdot (T_\mathrm{exp} - t_0)^{-\beta} + F_0,
\end{equation}
where $t_0$ and $F_0$ account for an offset in exposure time and flux
respectively. Our best-fit parameters are given in
Table~\ref{tab:sensitivity_parameters}. Since all \mbox{\textsl{HECATE}}
galaxies have $T_\mathrm{exp} \ge 75\,\mathrm{s}$, all potential candidates with
fluxes ${\gtrsim}10^{-13}\,\mathrm{erg}\,\mathrm{s}^{-1}\,\mathrm{cm}^{-2}$
should have been detected. For the galaxies located in areas of the sky with longer
$T_{\mathrm{exp}}$, the limiting flux is
${\sim}10^{-14}\,\mathrm{erg}\,\mathrm{s}^{-1}\,\mathrm{cm}^{-2}$. 
 
\begin{figure*}
  \sidecaption
  \includegraphics[width=12cm]{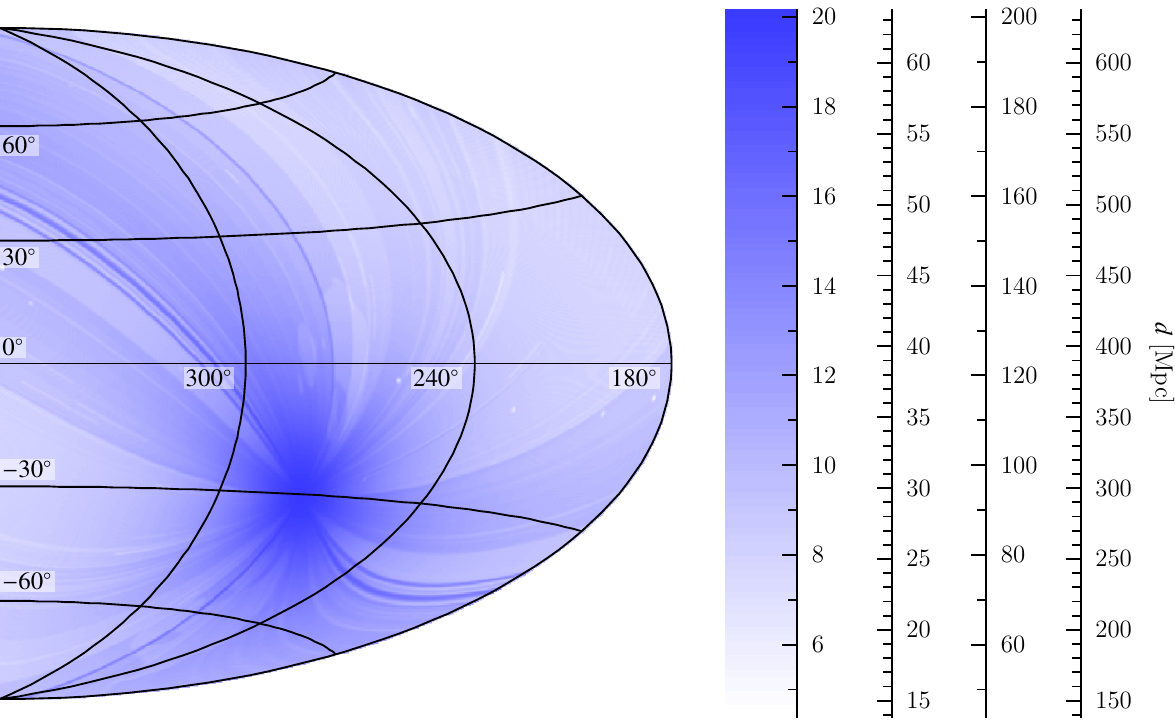}
  \caption{Spatially resolved map of maximum distances up to which a putative
    ULX with a given minimum luminosity would be detected in the eRASS1 with a
    detection likelihood of at least 20. The four different legends apply to
    minimum luminosities of
    $10^{39}\,\mathrm{erg}\,\mathrm{s}^{-1}$,
    $10^{40}\,\mathrm{erg}\,\mathrm{s}^{-1}$,
    $10^{41}\,\mathrm{erg}\,\mathrm{s}^{-1}$, and
    $10^{42}\,\mathrm{erg}\,\mathrm{s}^{-1}$,
  from left to right. See Table~\ref{tab:volumes} for numerical values of the volumes. Note that most \mbox{\textsl{HECATE}} galaxies are located within 200\,Mpc.
  }
  \label{fig:sensitivity_maps}
\end{figure*}
Using Eq.~\eqref{eq:sensitivity}, we converted the flux limit for each
\mbox{\textsl{HECATE}} galaxy into a point source luminosity limit. In
Fig.~\ref{fig:sensitivity_maps} we illustrate the maximum distance at which a
ULX is detectable as a function of its position on the sky. Our result implies
that for hyperluminous X-ray sources (HLX), with
$L_{\mathrm{X}}\ge10^{41}\,\mathrm{erg}\,\mathrm{s}^{-1}$, our main candidate
sample is complete out to ${\sim}70\,\mathrm{Mpc}$, while the ULX sample is
complete within ${\sim}7$\,Mpc. In order to estimate the luminosity dependent
volume in which eRASS1 is complete, we subdivided the sky into $N_\mathrm{pix}$
pixels of equal area using the Hierarchical Equal Area isoLatitude Pixelization
\citep[HEALPix;][]{gorski2005} package. For each pixel we then used
Eq.~\ref{eq:sensitivity} to determine the minimum flux necessary for a
detection, $F_\mathrm{lim}$, and estimated the total sensitive volume down to a
threshold luminosity, $L_\mathrm{lim}$,
\begin{equation}
  V_\mathrm{tot}(L_\mathrm{lim}) = \sum_{i = 1}^{N_\mathrm{pix}} \frac{1}{N_\mathrm{pix}}
    \frac{4}{3} \pi d_{i,\mathrm{lim}}^3 =
    \sum_{i = 1}^{N_\mathrm{pix}} \frac{1}{N_\mathrm{pix}} \frac{4}{3} \pi
    \left(
      \frac{L_\mathrm{lim}}{4 \pi F_{i,\mathrm{lim}}}
      \right)^{3/2} ,
  \label{eq:volume1}
\end{equation}
where $d_\mathrm{lim}$ is the maximum distance for each individual pixel.
Evaluating the sum yields
\begin{equation}
  V_\mathrm{tot} (L_\mathrm{lim}) = 2937\,\mathrm{Mpc}^3 \left(
    \frac{L_{\mathrm{lim}}}{10^{39}\,\mathrm{erg}\,\mathrm{s}^{-1}}
    \right)^{3/2}\quad.
  \label{eq:volume2}
\end{equation}
Table~\ref{tab:volumes} lists the total volumes and the number of galaxies to
which eRASS1 is sensitive for various luminosity thresholds.

\begin{table}
  \centering
  \caption{Sensitive volume for different lower luminosity thresholds.}
  \label{tab:volumes}
  \begin{tabular}{lrrr}
    \hline\hline
    $\log L_{\mathrm{lim}}\,[\mathrm{erg}\,\mathrm{s}^{-1}]$ &
    $V_{\mathrm{tot}}\,[\mathrm{Mpc}^3]$ & $N_{\mathrm{gal}}$ &
    $f_{\mathrm{gal}}$ \\
    \hline
  39 & $3 \times 10^3$ & 1750 & 1.7\% \\
    40 & $93 \times 10^3$ & 8033 & 7.8\% \\
    41 & $3 \times 10^6$ & 39979 & 38.8\% \\
    42 & $93 \times 10^6$ & 103069 & 100\% \\
    \hline
  \end{tabular}
  \tablefoot{The limits were calculated using Eq.~\ref{eq:volume2} with
  $N_\mathrm{pix}=199608$. We also list the total and relative number of
  galaxies in \mbox{\textsl{HECATE}} above each luminosity threshold. Down to
  $10^{42}\,\mathrm{erg}\,\mathrm{s}^{-1}$ eRASS1 is sensitive to all but 26
  galaxies.}
\end{table}

\subsection{Estimation of the AGN contamination}
\label{sec:contamination}
The majority of X-ray point sources detected at high Galactic latitudes are
expected to be background AGN \citep{predehl2021}. A fraction of these
appear within the $D_{25}$ regions of galaxies, but are actually located behind
them. If the AGN are bright enough, they may be misidentified as ULX. Since a
fraction of these AGN are not flagged by our screening
(Sect.~\ref{sec:identification}), they remain in the sample as interlopers
and skew further analyses. In this section we estimate this AGN contamination.

\citet{salvato2025} utilize a machine learning approach to identify optical and
infrared counterparts of eRASS1 sources based on training samples not
representative for ULX candidates, as they are X-ray point sources embedded in the
optical disks of galaxies, rendering their photometric redshift estimates
unreliable. Sources in the $D_{25}$ regions therefore need to be investigated
individually.

The sensitive flux range of eROSITA is fully covered by the well-established AGN
$\log N$-$\log S$ \citep{gilli2007}\footnote{
Retrieved from \url{http://www.bo.astro.it/~gilli/counts.html}
with
$F_{\mathrm{min}} = 10^{-15}\,\mathrm{erg}\,\mathrm{s}^{-1}$,
$F_{\mathrm{max}} = 10^{-10}\,\mathrm{erg}\,\mathrm{s}^{-1}$,
$N_{\mathrm{H},\mathrm{min}} = 10^{20}\,\mathrm{cm}^{-2}$,
$N_{\mathrm{H},\mathrm{max}} = 10^{26}\,\mathrm{cm}^{-2}$,
$z_{\mathrm{min}} = 0$,
$z_{\mathrm{max}} = 10$,
$L_{\mathrm{min}} = 10^{35}\,\mathrm{erg}\,\mathrm{s}^{-1}$,
$L_{\mathrm{max}} = 10^{50}\,\mathrm{erg}\,\mathrm{s}^{-1}$,
no high-$z$ decline
}.
Based on this $\log N$-$\log S$ distribution, we determined the number of AGN
that would (erroneously) be identified as ULX in a given galaxy's $D_{25}$
region. We integrated the 0.5--2.0\,keV AGN XLF over the flux range corresponding
to the luminosity range of ULX in each galaxy, properly taking into account each
galaxy's sensitivity limit (Eq.~\ref{eq:sensitivity}). We converted the fluxes
from the eROSITA band to the 0.5--2.0\,keV band used by \citet{gilli2007}, by
assuming the spectral shape of the AGN is an absorbed power law with
$\Gamma=1.7$ and $N_\mathrm{H}=10^{21}\,\mathrm{cm}^{-2}$. We determined the
number of AGN expected in each galaxy, considering its area and accounting for the
nuclear exclusion region, arriving at the total number of expected contaminants
by summing over all galaxies. Taking into account the number of background AGN
identified in the sample by catalog matching and removing them from the ULX
catalogs, our results imply 26 remaining AGN interlopers in the main sample, 38 in the
extended sample, and therefore 64 in the combined main and extended sample
(Table~\ref{tab:background_agn}).

\begin{table}
    \caption{Estimated contamination by AGN in the main and combined samples.}
    \label{tab:background_agn}
    \centering
    \small
    \begin{tabular}{cccc}
      \hline\hline
      Sample   & $N_\mathrm{ULX}$ & $N_{\mathrm{AGN}}$ & $f$ \\
      \hline
      both      & 350              & 64 & 18\% \\
      main    & $\phantom{0}$90  & 26 & 29\% \\
      \hline
    \end{tabular}
    \tablefoot{
      We determined the contamination by folding the flux-dependent
      probability for a detection with the XLF of \citet{gilli2007} and
      integrating over all galaxies. Because we identified three specific
      background sources in the samples, we removed this number from the values
      in this table.
    }
\end{table}

Our estimate of the contaminants solely relies on the geometry of the
galaxies and ignores the attenuation of background
AGN by matter in the galaxy. Using typical AGN spectra (power laws with $\Gamma
= 1.5$ to $\Gamma = 2.3$), absorption in the intervening galaxy can reduce the
0.2--8.0\,keV flux by ${\sim}66\%$, assuming a typical galaxy intrinsic
$N_{\mathrm{H}} = 10^{21}\,\mathrm{cm}^{-2}$ \citep[e.g.,][]{kaaret2017}. Our
estimate of AGN background contaminants can therefore be considered a
conservative upper limit.

\begin{figure*}
    \centering
    \includegraphics[width=17cm]{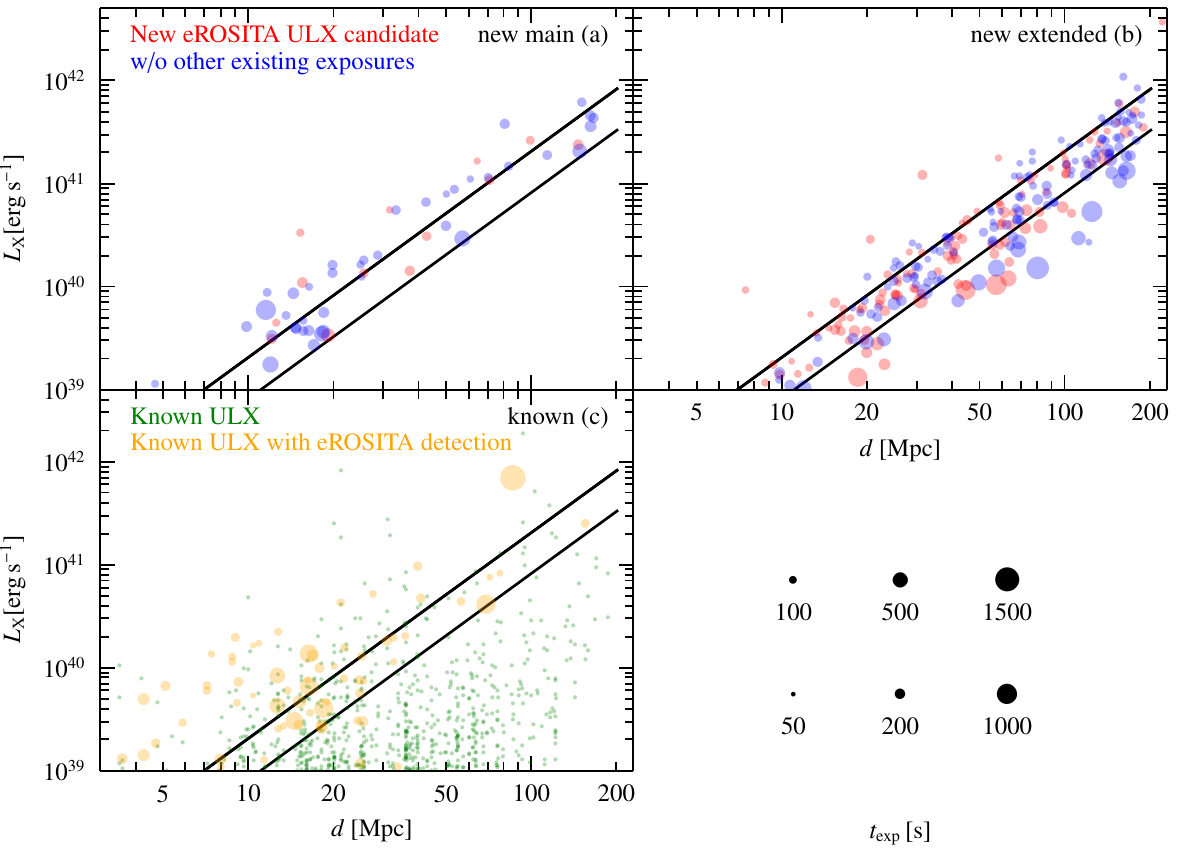}
    \caption{Luminosity distribution with respect to distance of the main new
    (a), extended new (b), and known (c) sample. We representatively show the
    sample by \citet{walton2022} as known sources without eROSITA detection and
    selected the highest available luminosity for each source from their catalog.
    The luminosities of known sources with eROSITA detection were determined
    using data from eRASS1. We converted these luminosities to the same energy band as
    used for eROSITA. The solid black lines denote the detection limits for the
    exposure at the ecliptic and the median exposure. The detection limit at the
    ecliptic pole is not covered by the diagram. Sources within the central
    exclusion zone of their hosts are not shown in the third panel. We have omitted
    uncertainties in the diagrams, as they would visually interfere with sizes of the
    circles indicating the exposure time. The average luminosity uncertainty in
    the first panel is ${\sim}58\%$, in the second panel it is ${\sim}77\%$, and of all
    eROSITA measurements in the third panel it is ${\sim}63\%$.}
    \label{fig:lumdist}
\end{figure*}

We identified 41\% of the sources of our main sample in other published catalogs
\citep{kovlakas2020,walton2022,bernadich2022,tranin2024}, while in the extended
sample this fraction is only 6\%. This illustrates the higher fraction of
contaminants in the extended sample, given its lower detection likelihood limit.
By comparing the number of expected background AGN in both samples
(Table~\ref{tab:background_agn}) we make a similar observation. With the
availability of eRASS:5 data and dedicated follow-up observations the validity of
some of the sources in the extended sample will be confirmed in the future.

\subsection{Source confusion}\label{sec:sourceconfusion}
With increasing distance, the limited angular resolution of eROSITA causes
clustered sources and extended regions of emission, such as star-forming
regions, to be detected as single, spuriously luminous point sources. This
confusion with point sources can add interlopers to the identified ULX
population and skew population analyses. Compared to other samples assembled
by \textsl{XMM-Newton}, \textsl{Chandra},
or \textsl{Swift}, which tend to have a better spatial resolution, this effect is more
relevant at smaller distances for our sample.

Star-forming regions are typically at most ${\sim}0.5\,\mathrm{kpc}$ in size
\citep{anastasopoulou2016,kovlakas2020}. In the full eRASS1 catalog we found the
smallest extended sources to be detected with an extent of ${\sim}8''$,
translating to a maximum distance of ${\sim}13\,\mathrm{Mpc}$ beyond which such
structures cannot be distinguished from point sources. This particularly
includes the XRB population, rendering them apparent ULXs with artificially
inflated luminosities. Of our main sample, only 30 sources reside closer than
13\,Mpc, rendering more detailed population studies focusing on this subsample
statistically challenging.

Due to the anticorrelation between the angular extent and distance of
galaxies, combined with the increasing luminosity required for the detection of
a point source, galaxies with smaller extent sample only the upper end of the
ULX XLF. As a result, in galaxies affected more by source confusion, only the
most luminous ULXs and assemblies of confused sources can be detected. To
investigate the degree of extended emission regions confused as single sources
in our sample we compared their luminosity with the integrated luminosity
expected from the gaseous emission of their host galaxy based on standard
$L_{\mathrm{X}}$-SFR scaling relations.

\begin{figure}
  \resizebox{\hsize}{!}{\includegraphics{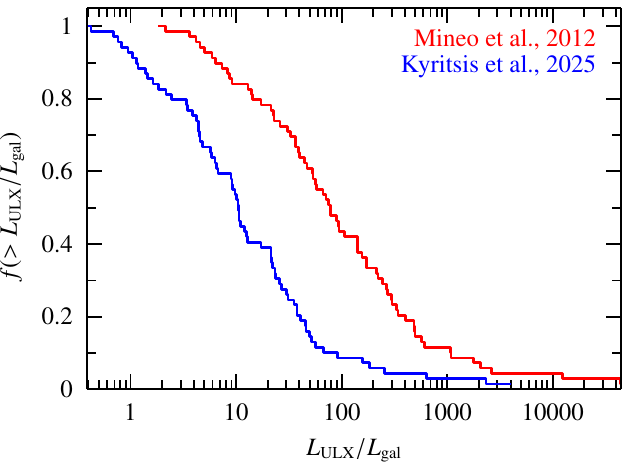}}
  \caption{Cumulative distribution of the ratio between the ULX luminosity
  $L_{\mathrm{ULX}}$ and the total luminosity $L_{\mathrm{gal}}$ of their
  respective host galaxies expected from 0.5--2\,keV scaling relations for late-type
  galaxies \citep{mineo2012a,kyritsis2025}.}
  \label{fig:luminosity_ratio}
\end{figure}
We used the scaling relation of \citet[][their Eq.~1]{mineo2012a},
$L_{\mathrm{X}}/\mathrm{SFR} \simeq 8.3 \times
10^{38}\,(\mathrm{erg}\,\mathrm{s}^{-1})/(M_{\odot}\,\mathrm{yr}^{-1})$, to
calculate the total luminosity expected in the 0.5--2.0\,keV band from the
gaseous emission in all late-type ULX host galaxies. After converting the
luminosities of all candidates to the same energy band, we calculated for each
galaxy the ratio between the candidate luminosity and that expected from the
total respective SFR.

In Fig.~\ref{fig:luminosity_ratio} we show the cumulative distribution of the
resulting ratios. For ${\sim}85\%$ of the galaxies, the luminosity of the ULX
candidate exceeds the emission expected from the SFR by at least 1\,dex, and for
${\sim}40\%$, it is exceeded by at least 2\,dex, suggesting it is unlikely that our candidates
are the result of hot gas in the hosts confused as point sources.
However, the confusion of multiple luminous XRBs in the host cannot be
ruled out by this method. We therefore used the recently obtained scaling
relation by \citet{kyritsis2025},
$\log(L_{\mathrm{X}}/\mathrm{erg}\,\mathrm{s}^{-1}) = 0.85
\log(\mathrm{SFR}/\mathrm{M}_{\odot}\,\mathrm{yr}^{-1}) + 39.67$, which also
includes the emission of resolved and unresolved XRBs, to perform the same
analysis, also displayed in Fig.~\ref{fig:luminosity_ratio}. At least
${\sim}95\%$ of candidates exceed the luminosity expected from their hosts,
and ${\sim}40\%$ by at least 1\,dex. With the area used for the
extraction of the candidates only being a fraction of that of the entire galaxy,
this reinforces that the candidates are likely not a result of confused
emission.

Early-type galaxies harbor point sources in the form of low mass XRBs (LMXBs)
but overall have a smooth distribution of their X-ray and optical emission due
to their old stellar population. We avoided the selection of such galaxies by
ignoring all sources flagged as extended in the eRASS1. AGN outflows or galactic
superwinds may form diffuse emission clumps that can be confused with point
sources. We therefore tested the hosts of all 13 candidates of our main sample
identified in early-type galaxies for the presence of AGN. None of these hosts
are listed in the AGN catalog of \citet{veron-cetty2010}, while
\mbox{\textsl{HECATE}} classifies NGC\,1439 and NGC\,4762 as AGN. Only one of
the galaxies, NGC\,2235, has an eRASS1 source listed at its center. The ULX in
this galaxy was also identified by \citet{walton2022}. We therefore consider
confusion with gaseous structures unlikely for the majority of our sample.
As for the candidates in late-type galaxies, confusion of multiple point sources
cannot be ruled out by this method.

\section{The eROSITA ULX sample in the context of previous searches}\label{sec:discussion}

\begin{table*}
  \centering
  \caption{Comparison between the fundamental properties of the ULX catalog
    derived in this work and past studies.}
  \label{tab:studies}
  \begin{tabular}{lccccc}
    \hline
    \hline
     Work\tablefootmark{a} & This work & T24 & W22 & B22 & K20 \\
    \hline
    Base catalog(s)\tablefootmark{b} & eRASS1 & 4XMM-DR11/CSC2/2SXPS & 4XMM-DR10/CSC2/2SXPS & 4XMM-DR9 & CSC2 \\
    Energy range\tablefootmark{c} [keV] & 0.2--5 & 0.2--12/0.5--7/0.3--10 & 0.2--12/0.5--8.0\tablefootmark{1}/0.3-10 & 0.2--12 & 0.5--8 \\
    Nuclear zone\tablefootmark{d} [$''$] & 12 & ${>}3$ & 9/6.1/9 & ${>}3$ & 3 \\
    Galaxy catalog\tablefootmark{e} & \mbox{\textsl{HECATE}} & \mbox{\textsl{GLADE}} & \mbox{\textsl{HyperLEDA}} \& CNG & \mbox{\textsl{HECATE}} & \mbox{\textsl{HECATE}} \\
    Search area\tablefootmark{f} & $D_{25}$ & $D_{\mathrm{Holm}} \simeq 1.26 D_{25}$ & $D_{25}$ & $D_{25}$ & $D_{25}$ \\
    Total candidates\tablefootmark{g} & 90 & $1901+191$\tablefootmark{2} & 1843 & 779 & 629 \\
    Host galaxies\tablefootmark{h} & 85 & 1303 & 951 & 517 & 309 \\
    Contamination\tablefootmark{i} [\%] & ${\lesssim} 29$ & ${\lesssim} 2$ & ${\sim} 20$ & ${\sim} 2$ & ${\sim} 20$ \\
    Completeness\tablefootmark{j} [Mpc] & ${\sim} 7$ & n/a & n/a & 29 & 40 \\
    \hline
  \end{tabular}
  \tablefoot{
    \tablefoottext{a}{The publications are abbreviated as T24: \citet{tranin2024}; W22: \citet{walton2022}; B22: \citet{bernadich2022}; K20: \citet{kovlakas2020}}
    \tablefoottext{b}{X-ray source catalog(s) the candidate selection was performed on}
    \tablefoottext{c}{Energy range in which the stated fluxes were determined}
    \tablefoottext{d}{extent of the nuclear exclusion region, variable size for T24 and B22}
    \tablefoottext{e}{Galaxy catalog(s) used for the host identification}
    \tablefoottext{f}{Parameter used for the optical extent of the galaxies}
    \tablefoottext{g}{Total number of ULX candidates}
    \tablefoottext{h}{Number of galaxies hosting the candidates}
    \tablefoottext{i}{Fraction of unknown contaminants}
    \tablefoottext{j}{Distance limit up to which the catalog is complete in X-rays, ignoring the completeness of galaxies}
    \tablefoottext{1}{Energy range for the ACIS. For the HRC it is 0.2--10\,keV}
    \tablefoottext{2}{\citet{tranin2024} state the number of ULXs and HLXs separately}
  }
\end{table*}

In recent years the combined archival data of current X-ray observatories was
used to identify ${\sim}2000$ ULX candidates
\citep{kovlakas2020,bernadich2022,walton2022,tranin2024}. The necessary
observations of galaxies by these instruments were generally carried out for
unrelated scientific rationales. While offering substantial depth for each
observation, the number of observed galaxies is limited and their selection
highly biased. Utilizing data from eRASS1, we could mitigate the biased selection
by trading against overall depth. In Table~\ref{tab:studies} we provide the key
properties of our catalog and contrast them with these other recent samples.

The detectability of galaxies generally correlates with their optical flux and
therefore decreases with distance. Since eROSITA is, down to
$10^{39}\,\mathrm{erg}\,\mathrm{s}^{-1}$, mostly sensitive to galaxies in the
local Universe, the choice of the host galaxy catalog does not influence the
overall result, as long as they equally cover this local range, which is the
case for all catalogs listed in Table~\ref{tab:studies}. We choose
\textsl{HECATE} because it offers the most robust distance estimates, which
directly enters the calculation of the ULX candidate luminosities, as well as
robust estimates for SFR, stellar mass and metallicity, enabling further studies
between the ULX population and their hosting environment (see paper~II).

The completeness of X-ray selected catalogs is often expressed as the area
covered down to a limiting flux. The completeness of ULX catalogs is a
combination of that of the host galaxy sample and that providing the X-ray
detections, and is therefore difficult to derive. Additionally, the choices for
the extent of galaxies and the nuclear exclusion zone influences the
completeness of ULX catalogs. The nature of ULXs provides a natural lower
luminosity threshold. Therefore ULX studies typically state the farthest
distance up to which all sources down to this luminosity are detected. This
ignores the fact that pencil beam surveys are categorically insensitive to most
galaxies.

\begin{figure}
  \resizebox{\hsize}{!}{\includegraphics{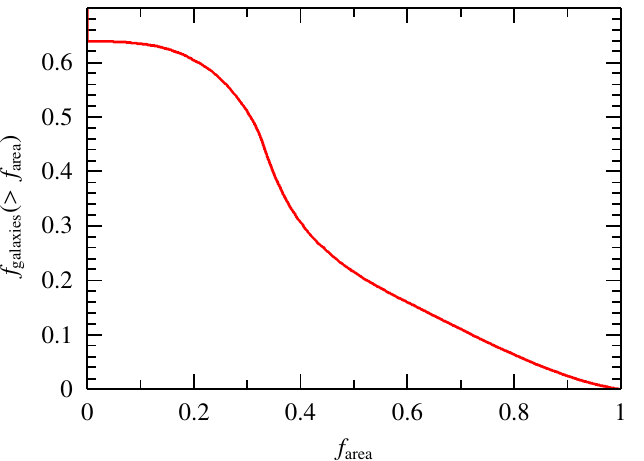}}
  \caption{
    Cumulative distribution of the fractional area covered by eROSITA in each
    galaxy. For ${\sim}36\%$ of the galaxies, $f_{\mathrm{area}} = 0$.
  }\label{fig:hecate_areas}
\end{figure}

With the eROSITA survey this aspect is drastically different. In principle
eROSITA is sensitive to all \mbox{\textsl{HECATE}} galaxies in the footprint of eRASS1, which amounts
to ${\sim}10^5$ objects. However, to avoid the inclusion of central AGN, an area around
the core of the galaxy is typically excluded from the search. Given the coarser
spatial resolution of eROSITA, we require this region to be larger compared to
past studies. As a result, with increasing distance a larger fraction of
galaxies is fully engulfed by this exclusion zone, decreasing the effective
fractional area $f_{\mathrm{area}}$ sensitive to our ULX search. In
Fig.~\ref{fig:hecate_areas} we show the cumulative distribution of
$f_{\mathrm{area}}$ in each galaxy. For ${\sim}36\%$ of all observed galaxies
$f_{\mathrm{area}} = 0$, for ${\sim}22\%$ at least half of the area is covered.

\begin{figure}
  \resizebox{\hsize}{!}{\includegraphics{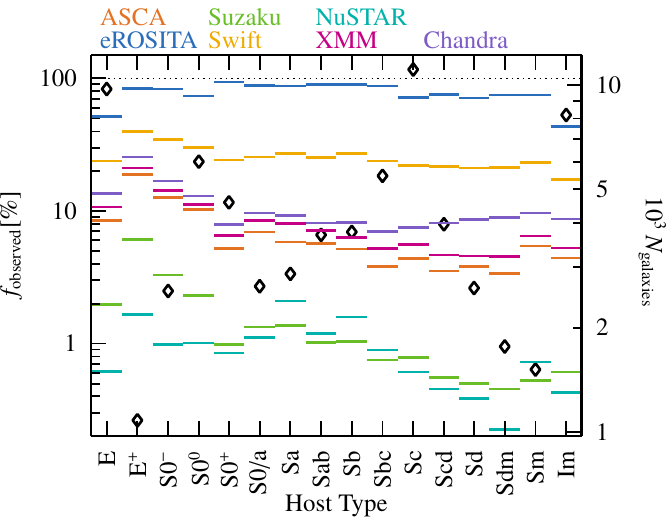}}
  \caption{Fraction of galaxies on the western Galactic hemisphere listed in
    \mbox{\textsl{HECATE}} observed by other X-ray instruments to any degree before the end of
    eRASS1. Galaxies without a morphological type in \mbox{\textsl{HECATE}} have been
    omitted. For eROSITA only galaxies for which $r_1$ exceeds the nuclear
    exclusion zone are considered. The diamonds show the total number of
    \mbox{\textsl{HECATE}} galaxies for each type with the axis on the right
    hand side.
    }\label{fig:hecate_exposures}
\end{figure}

In Fig.~\ref{fig:hecate_exposures} we compare the number of galaxies in the
western Galactic hemisphere observed by past and current X-ray observatories to
the number of galaxies for which eROSITA has $f_{\mathrm{area}} > 0$, resolved
by their morphological type. We limit ourselves to the ${\sim}65\%$ of galaxies for which
\mbox{\textsl{HECATE}} provides morphological information. Overall, eROSITA provides the best
coverage among all types of galaxies, apart from irregulars and ellipticals
exceeding 70\%, up to more than 90\%. Interestingly, for elliptical galaxies the
fraction is smaller than for other types of galaxies. This indicates that
ellipticals generally have a smaller apparent extent, which is likely related to
the selection function of \mbox{\textsl{HECATE}} favoring such galaxies at
farther distances. In contrast, the instruments utilized for recent ULX
catalogs, \textsl{Swift}, \textsl{XMM-Newton}, and \textsl{Chandra}, on average
cover only ${\sim}30\%$, ${\sim}10\%$, and ${\sim}10\%$ of the galaxies,
respectively, without accounting for the central exclusion zone utilized by
these other studies.

The main advantage for basing a ULX search on the eRASS1 is this unbiased
selection of galaxies. Nearby galaxies have larger extents and therefore larger
$f_{\mathrm{area}}$. Additionally, the survey is more sensitive at these closer
distances. Therefore, the majority of all galaxies within this local volume up
to ${\sim}7\,\mathrm{Mpc}$ is covered by our ULX search. In contrast, searches
based on pencil-beam surveys, resulting in deeper exposures and more ULX
candidates, generally provide farther completeness limits for their observed
galaxies, but the set of targeted galaxies is substantially lower and more
affected by selection biases.

\begin{figure}
  \resizebox{\hsize}{!}{\includegraphics{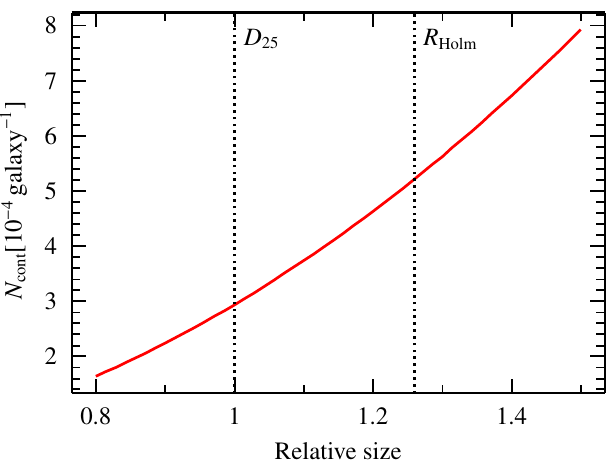}}
  \caption{
    Average number of expected background AGN contaminants per galaxy in the eROSITA 
    sample as a function of  the relative size of the ellipses defining the galaxy
    area. The $D_{25}$ ellipses are used as the reference, with their extent and
    that of the Holmberg radius also marked in the figure. We expect an increase
    of ${\sim}59\%$ in size when choosing the Holmberg radius instead of $D_{25}$.
  }
  \label{fig:contaminants_radius}
\end{figure}

The definition of the area of a galaxy directly influences the number of
candidates and AGN interlopers picked up during the selection procedure.
Generally, the density of ULXs follows the density of the stellar population in
a given galaxy \citep{wang2016,kovlakas2020,tranin2024}, which is peaked in the
center and drops toward the outer area. The choice of the central exclusion
zone therefore disproportionally influences the total number of ULXs a search
can recover. In contrast, the spatial density of AGN interlopers is constant.
Therefore, larger nuclear zones increase the fraction of AGN contaminants in the
ULX sample, leading to the observed anticorrelation between the nuclear zone and
number of candidates in Table~\ref{tab:studies} and the more severe
contamination in our sample in particular.

Similarly, larger extents of galaxies naturally accumulate more AGN interlopers
in the ULX sample. To illustrate this effect, we calculated the number of
interlopers per galaxy depending on their extent, relative to the $D_{25}$
regions. As shown in Fig.~\ref{fig:contaminants_radius}, choosing the Holmberg
radius instead of the $D_{25}$ isophote increases the contamination by
${\sim}59\%$. We therefore conclude that the drawback of the increased number of
AGN interlopers would outweigh the gain in additional candidates.

\section{Evidence of variability in ULXs}\label{sec:variability}
Dedicated observatories do not only provide much longer observation durations
compared to eROSITA, but especially \textsl{Swift} repeatedly captures data of
multiple targets over a longer time span. This observing pattern enables
variability studies
\citep[e.g.,][]{walton2022,bernadich2022,brightman2023,tranin2024}. In fact, the
variable nature of ULXs can often cause them not to appear as such.
\citet{walton2022} identified several sources that show peak luminosities above
the ULX threshold but are also been found to be less luminous in other
observations.

\begin{table}
\centering
\caption{
  Number of candidates in the main sample previously identified by other studies.
}
\label{tab:others}
\begin{tabular}{lcccc}
\hline
\hline
Work\tablefootmark{a} & \begin{tabular}{@{}c@{}}detection \\ expected\tablefootmark{b}\end{tabular} & \begin{tabular}{@{}c@{}}detected \\ (unique)\tablefootmark{c}\end{tabular} & \begin{tabular}{@{}c@{}}expected \& \\ detected\tablefootmark{d}\end{tabular} \\
\hline
T24 & 206 & 46 (23) & 43 (22)\\
W22 & 285 & 31 (29) & 29 (28)\\
B22 & 57 & 17 (15) & 11 (11)\\
K20 & 76 & 18 (16) & 15 (15)\\
\hline
\end{tabular}
\tablefoot{
\tablefoottext{a}{Publications are abbreviated as T24: \citet{tranin2024}; W22: \citet{walton2022}; B22: \citet{bernadich2022}; K20: \citet{kovlakas2020}}
\tablefoottext{b}{Number of sources that are expected to be detected, by
converting the fluxes to the eROSITA energy band, compared to the lower flux
limit determined based on Sect.~\ref{sec:completeness}, and considering the
$D_{25}$ regions and the nuclear exclusion zone}
\tablefoottext{c}{Number of sources in each work for which a counterpart in our
main sample was identified within a radius of $8''$. eROSITA's
inferior spatial resolution can lead to the identification of multiple
candidates, and therefore the number of unique eROSITA matches is given in
parenthesis}
\tablefoottext{d}{Number of sources that were expected to be detected and were
identified}
}
\end{table}
To test for such changes in luminosity we converted the fluxes from past studies
to the eROSITA band and compared them to the sensitivity estimates given in
Sect.~\ref{sec:completeness}. We thus derived subsamples that we expected to be
also present in our sample. For the sample by \citet{walton2022} we selected the
peak fluxes and used the energy band of the respective instrument. We list the
resulting identifications in Table~\ref{tab:others}, which indicates that we
only identified a small fraction of the expected sources.

Thirteen of the 90 candidates in the main sample have not been previously
identified as ULX but were observed before the end of eRASS1 by at least one of
\textsl{ASCA} ($25'$), \textsl{Chandra} ($15'$), \mbox{\textsl{NuSTAR}} ($6'$),
\textsl{Suzaku} ($9'$), \textsl{Swift} ($12'$), or \textsl{XMM-Newton} ($15'$),
with assumed fields of view in parentheses, possibly indicating strong luminosity
variation, given the better sensitivity of these missions. In
Fig.~\ref{fig:lumdist} we mark such sources as well as previously known
candidates that we expected to identify, given their flux reported in the
literature. Most of the known undetected candidates tend to be far below the
median sensitivity line, and hence they would not have registered in eRASS1 if they maintained
their luminosity. Undetected ULXs with reported fluxes near our sensitivity
limit must have faded into the background. This behavior shows that ULXs, in
general, can be characterized as highly variable on the timescale of months or
years. Even though the physics of supercritical accretion disks is still poorly
understood, this timescale aligns with observations of LMXBs driven by disk
instabilities \citep{dubus2001,lasota2001,heinke2025}, and type~II outbursts of
HMXBs fueled by changes in the circumstellar disk of the donating Be star
\citep{reig2011,neumann2023}. The XLF of LMXBs shows a break below the ULX
threshold \citep{sarazin2000,kim2004,humphrey2008,kim2010,wang2016}, suggesting
that HMXBs could be the dominant contribution of binary systems to
the ULX population. This suggestion aligns with the observed variability and the detection of
pulsations from ULXs \citep{bachetti2014,israel2017a}, while some of the more
persistent sources could still be powered by an IMBH. Strong variability was
also found in other known ULXs
\citep{motch2014b,walton2015,brightman2016,earnshaw2020,earnshaw2022},
potentially as a result of the propeller effect
\citep{tsygankov2016,earnshaw2020,middleton2023}, tying into the picture of a
sizeable portion of the ULX population being powered by NSs. Alternative
explanations proposed for the long superorbital periods observed in some ULXs
involve the precession of the NS itself \citep{vasilopoulos2020} or of
structures within the accretion flow, such as a warped disk \citep{motch2014b}
or outflowing material \citep{middleton2018,middleton2019,khan2022}.

eROSITA performed multiple observations of each source, limited to
${\sim}40\,\mathrm{s}$, separated by 4\,h intervals and repeated
between 6 and roughly 1000 times at the survey equator and pole respectively. In
our main sample the candidate with the shortest $T_\mathrm{exp}$ was observed
for only ${\sim}130\,\mathrm{s}$, while the most observed one has
$T_\mathrm{exp} {\sim}3300\,\mathrm{s}$. For our main sample, the median number
of total detected counts is 20, which does not permit any further variability
study. In future studies data from all five available eROSITA surveys (eRASS1
through eRASS5) can be used to constrain the variability of the candidates on a
half-year cadence.

\section{Summary and outlook}\label{sec:summary}
We used the main eRASS1 catalog (Sect.~\ref{sec:erosurvey}) to identify ULXs in
the $D_{25}$ ellipses of \mbox{\textsl{HECATE}} galaxies
(Sect.~\ref{sec:hecate}). By ignoring all sources closer than $12''$ to the core
of the $D_{25}$ regions, we avoided the selection of AGN. Using extracted spectra,
we calculated the flux of each candidate in Sect.~\ref{sec:identification}, and
using the distance of the putative host galaxy, we calculated the luminosity. To identify
other known contaminants, we matched our list of candidates to external databases
and catalogs, which let us remove supernovae, AGN, and stars
(Sect.~\ref{sec:cleaning}). Inspecting all initial candidates manually led us to
two samples of ULX candidates. With 90 candidates, the main sample consists of
sources with a detection likelihood above 20, and this sample can be used in further
population studies. We also identified 260 candidates with detection likelihoods
between 10 and 20, which are likely contaminated by significant amounts of
spurious detections.

To characterize the sensitivity of our selection scheme, we performed a
simulation of eRASS1 based on actual eROSITA attitude data, followed by
source detection identical to that used for the main eRASS1 catalog
(Sect.~\ref{sec:completeness}). The resulting flux limit indicates that our ULX
catalog is complete up to $d \sim 7\,\mathrm{Mpc}$, while we expect all HLXs in
known galaxies within 70\,Mpc to be detected, if they exceed the flux limit at
the time of observation. The simulation, combined with the $\log N$-$\log S$ of
AGN, allowed us to constrain the number of unidentified contaminants from
background AGN to at most ${\sim}29\%$ in our main sample
(Sect.~\ref{sec:contamination}). By comparing the angular extent and distances
of the observed galaxies to the spatial resolution of eROSITA, we derived a
distance of $d \sim 13\,\mathrm{Mpc}$ for the onset of source confusion.
We also considered the properties of the hosting galaxies and find this issue unlikely to be relevant for the main sample
(Sect.~\ref{sec:sourceconfusion}).
Interestingly, we also find that a significant number of already known ULXs are
absent from our sample, even though their reported fluxes are well within the
eROSITA sensitivity range. We interpret the absence of these sources as an
indication of variability being a common property among ULXs. The archival data from pointed observations and the rich multisurvey data
obtained by eROSITA will be useful for performing a more systematic analysis of this variability.

We have identified 53 new ULX candidates in our main sample and 245 in the
extended sample in other recent ULX catalogs. With the indiscriminate selection
of galaxies distributed over an entire hemisphere, our catalog
is unaffected by biases regarding properties of the host galaxies, such as their
morphological type. This is in contrast to previous studies that selected
candidates from pointed or limited slew observations. We have also identified sources
as ULXs that were observed before and not classified as such, and, vice versa, we
did not identify several candidates expected from other studies, indicating
variable behavior over longer timescales (Sect.~\ref{sec:variability}).

While some ULXs have unambiguously been shown to be NSs in binary systems
\citep{bachetti2014,israel2017}, the nature of many other sources is still
unclear. The high luminosity well beyond the Eddington limit of a stellar mass
BH suggests the possibility that a subpopulation of ULXs might comprise
accreting IMBHs \citep{webb2012, mezcua2015}. If the ULX population can be
broken down into several categories, a break in the luminosity function could
reveal the separation of these populations. Such studies in particular require
complete samples of ULXs and galaxies cleaned from contaminants and assembled in
an unbiased manner. With our catalog based on eROSITA data, we have laid the
groundwork for further systematic studies of the ULX population in the local
Universe. In the accompanying paper~II, we use our main sample and sensitivity
and contamination estimates to investigate the luminosity function and relations
to hosting galaxies while taking into account and correcting the effects of
incompleteness and contamination.

In future works, harvesting data from eRASS1 through eRASS5 will increase the
unbiased ULX sample detected by eROSITA. Using the survey's exposure, we estimate
that a ULX sample selected from eRASS:5 should be complete within
${\sim}14\,\mathrm{Mpc}$. The half-year cadence of the individual surveys will
also enable systematic variability studies of ULXs and thus shed light on the
possible presence of orbital phases or spectral state transitions typical for
XRBs.

\section*{Data availability}
The main and extended catalog, as well as the list of identified contaminants,
are available in the online version of the journal, the eROSITA DR1 webpage at
\url{https://erosita.mpe.mpg.de/dr1/AllSkySurveyData_dr1/Catalogues_dr1/}, and
at the CDS via anonymous ftp to \url{cdsarc.u-strasbg.fr}
(\texttt{130.79.128.5}) or via
\url{http://cdsweb.u-strasbg.fr/cgi-bin/qcat?J/A+A/XXX/XXX}.

\begin{acknowledgements}
We thank the anonymous referee for the helpful and insightful comments which
improved the quality of this article.

We acknowledge funding from Deutsches Zentrum f\"ur Luft- und Raumfahrt (DLR) grants 50\,OR\,2309, 50\,QR\,2103, and 50\,QR\,2503, and from Deutsche Forschungsgemeinschaft grant 414059771.
S.H.\ was partly supported by the German Science Foundation (DFG grant numbers WI 1860/14-1 and 434448349).
LD acknowledges funding from German Research Foundation (DFG), Projektnummer 549824807.
TPR acknowledges support from the Science and Tech-
nology Facilities Council (STFC) as part of the consolidated grant
award ST/X001075/1. 

This work is based on data from eROSITA, the soft X-ray instrument aboard SRG, a joint Russian-German science mission supported by the Russian Space Agency (Roskosmos), in the interests of the Russian Academy of Sciences represented by its Space Research Institute (IKI), and the Deutsches Zentrum für Luft- und Raumfahrt (DLR). The SRG spacecraft was built by Lavochkin Association (NPOL) and its subcontractors, and is operated by NPOL with support from the Max Planck Institute for Extraterrestrial Physics (MPE). The development and construction of the eROSITA X-ray instrument was led by MPE, with contributions from the Dr.\ Karl Remeis Observatory Bamberg \& ECAP (FAU Erlangen-N\"urnberg), the University of Hamburg Observatory, the Leibniz Institute for Astrophysics Potsdam (AIP), and the Institute for Astronomy and Astrophysics of the University of Tübingen, with the support of DLR and the Max Planck Society. The Argelander Institute for Astronomy of the University of Bonn and the Ludwig-Maximilians-Universit\"at Munich also participated in the science preparation for eROSITA. The eROSITA data shown here were processed using the eSASS software system developed by the German eROSITA consortium.

This research has made use of ISIS functions (ISISscripts) provided by
ECAP/Remeis observatory and MIT (http://www.sternwarte.uni-erlangen.de/isis/).

This work has made use of data from the European Space Agency (ESA) mission Gaia
(https://www.cosmos.esa.int/gaia), processed by the Gaia Data Processing and
Analysis Consortium (DPAC, https://www.cosmos.esa.int/web/gaia/dpac/consortium).
Funding for the DPAC has been provided by national institutions, in particular
the institutions participating in the Gaia Multilateral Agreement.

This research has made use of the SIMBAD database,
operated at CDS, Strasbourg, France, and of the VizieR catalogue access tool, CDS,
Strasbourg, France (DOI: 10.26093/cds/vizier). The original description 
of the VizieR service was published in 2000, A\&AS 143, 23.
\end{acknowledgements}

\bibliographystyle{aa}
\bibliography{main}

\begin{appendix}
  \section{Screening flags}
  \label{app:flags}
  In the following we list the flags available during the manual screening
  process.
  \begin{itemize}
    \item Spurious detection: Indicates whether the detection of the X-ray source
      itself is spurious.
    \item Foreground star: Indicates that the X-ray detection was caused by either
      an X-ray bright foreground star in the Milky Way or optical loading.
    \item Optical Point source: Indicates that the X-ray detection is associated
      with an optical point source that could not be clearly identified as a
      foreground star
    \item Diffuse X-ray emission: Indicates that the  X-ray detection is embedded
      into an area of increased diffuse radiation
    \item Wrong $D_{25}$ region: Indicates that the $D_{25}$ region in the
      \mbox{\textsl{HECATE}} does not fit the optical structure of the hosting galaxy
    \item Nuclear source: Indicates that X-ray detection is associated with the
      core of a galaxy
    \item Star-forming region: Indicates that the X-ray detection is embedded in
      an area of potential star-forming activity as determined from the optical
      images of the hosting galaxy
    \item Background source: Indicates that the X-ray detection is likely not
      associated with the hosting galaxy but rather an object in the background
    \item Wrong morphological type: Indicates that the morphological type listed
      in the \mbox{\textsl{HECATE}} is likely wrong
    \item Interacting galaxy: Indicates that the galaxy is gravitationally
      interacting with at least one other galaxy
  \end{itemize}

\begin{table}[h!]
\caption{Amount of flags assigned during the manual screening at the 50\% and 100\% thresholds.}
\begin{tabular}{lccc}
\hline\hline
Min.\ detection likelihood & 6 & 10 & 20 \\
\hline
Spurious detection & 53; 15 & 38; 10 & 6; 0 \\ 
Background source & 34; 7 & 27; 6 & 16; 5 \\ 
Foreground star & 5; 1 & 7; 1 & 9; 2 \\ 
Star-forming region & 37; 11 & 36; 12 & 34; 13 \\ 
Diffuse X-ray emission & 44; 9 & 45; 11 & 36; 6 \\ 
Interacting galaxy & 5; 2 & 8; 3 & 9; 3 \\ 
Wrong $D_{25}$ region & 13; 1 & 13; 2 & 15; 3 \\ 
Nuclear source & 6; 0 & 8; 0 & 12; 1 \\ 
Wrong morphological type & 2; 0 & 3; 0 & 2; 0 \\ 
Optical point source & 14; 5 & 17; 7 & 18; 9 \\ 
\hline
\end{tabular}
\tablefoot{
The flag assignments are listed for different detection likelihood thresholds.
Sources with negative eROSITA quality flags have been removed before the
screening.
}
\end{table}

\section{Column description}\label{app:columns}
\subsection{Main and extended sample}
\noindent \verb|IAUNAME|: IAU Name of the source in the eRASS1 catalog \citet{merloni2024}

\noindent \verb|RAJ2000, DEJ2000|: Equatorial coordinates from the eRASS1 $[^\circ]$

\noindent \verb|screen_background{,_n}|: Number of times the source was flagged as background source and how often it was investigated

\noindent \verb|screen_diffuse{,_n}|: Number of times the source was flagged as diffuse region of emission and how often it was investigated

\noindent \verb|screen_interacting{,_n}|: Number of times the hosting galaxy was flagged as interacting with another galaxy and how often it was investigated

\noindent \verb|screen_nuclear{,_n}|: Number of times the source was flagged in the center nuclear region of the hosting galaxy and how often it was investigated

\noindent \verb|screen_optical{,_n}|: Number of times the source was flagged as counterpart to an optical source and how often it was investigated

\noindent \verb|screen_sfr{,_n}|: Number of times the source was flagged as associated with a star-forming region and how often it was investigated

\noindent \verb|screen_flag_spurious{,_n}|: Number of times the source was flagged as spurious detection and how often it was investigated

\noindent \verb|screen_star{,_n}|: Number of times the source was flagged as associated with a star and how often it was investigated

\noindent \verb|screen_wd25{,_n}|: Number of times the $D_{25}$ region of the galaxy was flagged as incorrect and how often it was investigated

\noindent \verb|screen_wmorph{,_n}|: Number of times the morphological type of the hosting galaxy was flagged as incorrect and how often it was investigated

\noindent \verb|host_{M,L}b|: Absolute $b$-band magnitude and luminosity of the hosting galaxy $[\mathrm{mag},L_{\odot}]$ calculated from HECATE data

\noindent \verb|erass1_{tstart,tstop}|: First/last moment the source was observed during eRASS1 [days since eroday 1, MJD 51544]

\noindent \verb|erass1_t|: Midpoint of all eRASS1 observations [days since eroday 1, MJD 51544]

\noindent \verb|erass1_dt|: Time difference between first and last observation during erass1 [days]

\noindent \verb|erass1_exposure_{0,_8,_9}|: Total exposure time during eRASS1 for the spectra of all TMs combined, TMs 1--4 \& 6, and TMs 5 \& 7

\noindent \verb|cumulative_angle_area{,_norm}|: Cumulative sensitive area for one quadrant of all HECATE galaxies combined and integrated from the center to the $D_{25}$ limit over the angle starting from the semimajor axis to the source [$\mathrm{arcsec}^2$] and normalized to the unit circle for each galaxy

\noindent \verb|cumulative_radial_area{,_norm}|: Cumulative sensitive area of all HECATE galaxies combined and integrated radially from the center to the source [$\mathrm{arcsec}^2$] and normalized to the unit circle for each galaxy

\noindent \verb|eccentricity|: Eccentricity of the $D_{25}$ region of the hosting galaxy

\noindent \verb|phi|: Position angle of the source with respect to the semimajor axis of the hosting galaxy

\noindent \verb|r_abs|: Absolute angular separation of the source to the galaxy center $[^\circ]$

\noindent \verb|r_gal|: Absolute angular separation between the core of the hosting galaxy and the $D_{25}$ ellipse in the direction of the source $[^\circ]$

\noindent \verb|r_norm|: Angular separation between the core of the hosting galaxy and the source normalized by the separation from the core to the $D_{25}$ ellipse

\noindent \verb|fmin{,_ulx}|: Minimum flux $[\mathrm{erg}\,\mathrm{s}^{-1}\,\mathrm{cm}^{-2}]$ and luminosity required for a detection of any source with a likelihood of at least 20 in the hosting galaxy and for a ULX detection

\noindent \verb|{,e_}counts|: eROSITA X-ray counts extracted from the spectrum and associated uncertainty calculated using the mechanism by \citet{gehrels1986} $[\mathrm{erg}\,\mathrm{s}^{-1}]$

\noindent \verb|ecf_{0,8,9}|: Energy conversion factor for all TMs, TMs 1--4 \&6, as well as TMs 5 \& 7, 0.2\,keV--5.0\,keV

\noindent \verb|{,e_,E_}flux|: 0.2\,keV--5.0\,keV band flux with lower and upper uncertainty $[\mathrm{erg}\,\mathrm{s}^{-1}\,\mathrm{cm}^{-2}]$

\noindent \verb|{,e_,E_}lum|: 0.2\,keV--5.0\,keV band luminosity with lower and upper uncertainty

\noindent \verb|{,e_}rate|: 0.2\,keV--5.0\,keV count rate determined from extracted spectra withuncertainties calculated using the method by \citet{gehrels1986}

\noindent \verb|{,e_,E_}Chandra_HR_{hm,hs,ms}|: Hardness ratios in the bands used by the CSC2 and the associated upper and lower uncertainties

\noindent \verb|{,e_,E_}Swift_HR{1,2}|: Hardness ratios in the bands used by the 2SXPS and the associated upper and lower uncertainties

\noindent \verb|{,e_,E_}XMM_HR{1,2,3,4}|: Hardness ratios in the bands used by the 4XMM DR13 and the associated upperand lower uncertainties

\noindent \verb|f_milliquas|: Flag indicating a match in the Milliquas catalog version 8 \citet{flesch2023}

\noindent \verb|milliquas_ids|: List of potential Milliquas counterparts as Name

\noindent \verb|f_sdss|: Flag indicating a match in the SDSS DR15 catalog \citet{aguado2019}

\noindent \verb|sdss_ids|: List of potential counterparts in the SDSS DR15

\noindent \verb|f_simbad_star|: Flag indicating a match with a star type object in the Simbad database

\noindent \verb|simbad_star_ids|: List of potential counterparts with type star in the Simbad database

\noindent \verb|f_walton|: Flag indicating a match in the catalog by \citet{walton2022}

\noindent \verb|f_bernadich|: Flag indicating a match in the catalog by \citet{bernadich2022}

\noindent \verb|bernadich_ids|: Counterpart in the catalog by \citet{bernadich2022} in the SRCID column

\noindent \verb|f_kovlakas|: Flag indicating a match in the catalog by \citet{kovlakas2020}

\noindent \verb|kovlakas_ids|: Counterpart in the catalog by \citet{kovlakas2020} in the name column

\noindent \verb|f_tranin_hlx|: Flag indicating a match in the HLX sample by \citet{tranin2024}

\noindent \verb|tranin_hlx_ids|: Counterpart in the HLX sample by \citet{tranin2024} in the iauname column

\noindent \verb|f_tranin_ulx|: Flag indicating a match in the ULX sample by \citet{tranin2024}

\noindent \verb|tranin_ulx_ids|: Counterpart in the ULX sample by \citet{tranin2024} in the iauname column

\noindent \verb|erass_*|: Column directly from the eRASS1 main catalog \citet{merloni2024}

\noindent \verb|hecate_{,e_}d|: Distances of the hosting galaxy [Mpc] and uncertainty (Kyritsis et al., 2026, accepted)

\noindent \verb|hecate_{ra,dec}|: Right ascension and declination of the hosting galaxy $[^\circ]$ (Kyritsis et al., 2026, accepted)

\noindent \verb|hecate_logm_star|: Base-10 logarithm of the stellar mass of the hosting galaxy $[\log_{10} M_{\odot}]$ (Kyritsis et al., 2026, accepted)

\noindent \verb|hecate_m_star_method|: Method used to determine the stellar mass of the hosting galaxy (Kyritsis et al., 2026, accepted)

\noindent \verb|hecate_logsfr|: Base-10 logarithm of the star formation rate of the hosting galaxy $[\log_{10} M_{\odot}\,\mathrm{yr}^{-1}]$ (Kyritsis et al., 2026, accepted)

\noindent \verb|hecate_sfr_method|: Method used to determine the star formation rate of the hosting galaxy (Kyritsis et al., 2026, accepted)

\noindent \verb|hecate_objname|: Common name of the hosting galaxy (Kyritsis et al., 2026, accepted)

\noindent \verb|hecate_r{1,2}|: Semi-major and semi-minor axis of the hosting galaxy [$'$] (Kyritsis et al., 2026, accepted)

\noindent \verb|hecate_pa|: Position angle of the $D_{25}$ region of the hosting galaxy (Kyritsis et al., 2026, accepted)

\noindent \verb|hecate_pgc|: PGC number of the hosting galaxy (Kyritsis et al., 2026, accepted)

\noindent \verb|hecate_{,e_}t|: Morphological type of the hosting galaxy and uncertainty (Kyritsis et al., 2026, accepted)

\subsection{Additional columns for contaminants}
\noindent \verb|f_manual|: Source manually flagged as contaminant

\noindent \verb|f_sn|: Flag indicating a match with a known supernova in the OSNE \citet{guillochon2017} or an object of type supernova in the NED

\noindent \verb|sn_ids|: List of potential supernova counterparts as prefname (NED) and name (OSNE)

\noindent \verb|f_veron|: Flag indicating a match in the catalog by \citet{veron-cetty2010}

\noindent \verb|veron_ids|: List of potential counterparts in the catalog by \citet{veron-cetty2010}

\noindent \verb|f_gaia|: Flag indicating a match with a finite parallax larger than 0 in the Gaia DR3 catalog

\noindent \verb|gaia_ids|: List of potential Gaia DR3 counterparts as \verb|source_id|

\noindent \verb|f_tycho|: Flag indicating a match in the Tycho-2 catalog \citep{hog2000}

\noindent \verb|tycho_ids|: List of potential counterparts in the Tycho-2 catalog

\noindent \verb|f_yale|: Flag indicating a match in the Yale Catalogue of Bright Stars \citet{hoffleit1995}

\noindent \verb|yale_ids|: List of potential counterparts in the Yale Catalogue of Bright Stars

\end{appendix}

\end{document}